\documentclass[aps,prb,twocolumn,superscriptaddress]{revtex4-2}
\usepackage{bm}
\usepackage{graphicx}
\usepackage{mathrsfs}
\usepackage{amsmath,amssymb}
\usepackage{braket}

\makeatletter
\newcommand{\figcaption}[1]{\def\@captype{figure}\caption{#1}}
\newcommand{\tblcaption}[1]{\def\@captype{table}\caption{#1}}
\makeatother

\begin{document}
\title{Exploration of optimal hyperfine transitions for spin-wave storage in $^{167}$Er$^{3+}$:Y$_2$SiO$_5$}

\author{K.\ Matsuura}
	\affiliation{Division of Applied Physics, Hokkaido University, N13 W8, Kitaku, Sapporo 060-8628, Japan}
\author{S.\ Yasui}
	\affiliation{Division of Applied Physics, Hokkaido University, N13 W8, Kitaku, Sapporo 060-8628, Japan}
\author{R.\ Kaji}
    \email{r-kaji@eng.hokudai.ac.jp}
	\affiliation{Division of Applied Physics, Hokkaido University, N13 W8, Kitaku, Sapporo 060-8628, Japan}
\author{H.\ Sasakura}
    \email{hirotaka@eng.hokudai.ac.jp}
	\affiliation{Division of Applied Physics, Hokkaido University, N13 W8, Kitaku, Sapporo 060-8628, Japan}
\author{T.\ Tawara}
\email{tawara.takehiko@nihon-u.ac.jp}
	\affiliation{College of Engineering, Nihon University, Koriyama, Fukushima 963-8642, Japan}		
\author{S.\ Adachi}
\email{adachi-s@eng.hokudai.ac.jp}
	\affiliation{Division of Applied Physics, Hokkaido University, N13 W8, Kitaku, Sapporo 060-8628, Japan}

\date{\today}

\begin{abstract}
The dependence of the magnetic fluctuations and the spin coherence time $T_2^{\rm hyp}$ of the lowest Stark states $^4I_{15/2}\ (Z_1)$ in $^{167}$Er$^{3+}$:Y$_2$SiO$_5$ under zero magnetic field on Er  concentration is numerically investigated in the range of 10 to 100 parts per million (ppm). We investigate two primary sources of magnetic fluctuation limiting spin coherence: a constant contribution from host Y nuclei and a concentration-dependent component from dipole-dipole interactions among Er ions. Due to these two components, the Er-concentration dependence of $T_2^{\rm hyp}$ at the zero first-order Zeeman (ZEFOZ) points saturates for crystals with Er concentration below 10 ppm and no extension of the $T_2^{\rm hyp}$ is expected without an external magnetic field. 
Under a magnetic field, the longest $T_2^{\rm hyp}$ at a particular ZEFOZ point is expected to be over 170 s (90 s) for site 1 (site 2), which is more than $10^4$ times longer than that at zero field for 10-ppm $^{167}$Er$^{3+}$:Y$_2$SiO$_5$. Remarkably, these optimal ZEFOZ points form striking geometric patterns: a line for site 1 and a plane for site 2. This trend, which is favorable for experiments, can be explained by the anisotropy of the effective spin Hamiltonian parameters. Finally, the tolerance of the ZEFOZ point at each site with the longest $T_2^{\rm hyp}$ against the errors in the applied magnetic field vector is evaluated.
\end{abstract}

\maketitle

\section{Introduction}
The development of quantum networks, which connect quantum computation nodes via optical fiber, is a major goal in quantum science~\cite{duan2001long,kimble2008quantum,pompili2021realization}. A key enabling component is the quantum memory (QM), capable of storing photonic quantum states for long durations~\cite{awschalom2018quantum,atature2018material}. While various physical systems are under investigation~\cite{wang2017single, wang2020single, ritter2012elementary, cho2016highly, lenef1996electronic, sukachev2017silicon}, rare-earth-ion-doped crystals are particularly promising due to their long coherence times~\cite{ham1997frequency,equall1994ultraslow}. Systems such as $^{151}$Eu$^{3+}$:YSO and Pr$^{3+}$:YSO have demonstrated remarkable spin coherence times, extending up to several hours~\cite{zhong2015optically,wang2025nuclear,heinze2014coherence}. However, a significant drawback limits their direct use in quantum networks: their optical transitions (at 580~nm and 606~nm, respectively) are incompatible with the low-loss telecommunication C-band used in existing fiber infrastructure.

Erbium-doped yttrium orthosilicate ($^{167}$Er$^{3+}$:Y$_2$SiO$_5$) emerges as a compelling solution to this challenge, as its optical transition at $\sim$1.53 $\mu$m offers native compatibility with telecom networks~\cite{macfarlane1997measurement,bottger2006optical}. The primary obstacle, however, is that Er$^{3+}$ is a Kramers ion. Unlike non-Kramers ions such as Eu$^{3+}$ and Pr$^{3+}$ (with effective spin $S=0$), its non-zero effective electron spin ($S=1/2$) couples strongly to magnetic noise from the host crystal~\cite{guillot2006hyperfine,mcauslan2012reducing}. Consequently, the reported spin coherence times for Er:YSO are significantly shorter (e.g., 380 $\mu$s at 0 T~\cite{rakonjac2020long} and 1.3 s at 7 T~\cite{rancic2018coherence}) than those of its non-Kramers counterparts. On the other hand, the strong interactions in Er$^{3+}$ provide a critical advantage in storage bandwidth. Non-Kramers ions like Eu and Pr have small nuclear spin splittings (typically tens of MHz), which fundamentally limits the bandwidth of an optical signal that can be mapped onto them for storage. Er$^{3+}$, in contrast, possesses large hyperfine splittings (sub-GHz to GHz), offering the potential for much broader bandwidth quantum memories.

A powerful strategy to overcome this limitation is to operate at Zero First-Order Zeeman (ZEFOZ) transitions---specific ``sweet spots" where the qubit transition becomes insensitive to magnetic field fluctuations. This approach has proven successful in the simpler Kramers ion system, $^{171}$Yb$^{3+}$:YSO ($S=1/2$, nuclear spin $I=1/2$)~\cite{ortu2018simultaneous}. However, $^{167}$Er$^{3+}$ ($S=1/2$ and $I=7/2$) presents a greater challenge due to its complex 16-state hyperfine manifold. The strong state mixing induced by the hyperfine (HFI) and nuclear quadrupole (NQI) interactions creates a complex energy level structure, making the identification of optimal ZEFOZ points a non-trivial task.

This work confronts this challenge through a systematic numerical investigation. To provide a roadmap for extending the electron spin coherence time ($T_2^{\rm hyp}$) via spin-wave storage~\cite{afzelius2010demonstration,minavr2010spin,businger2020optical,stuart2024progress}, we model the magnetic fluctuations arising from two distinct sources: the nuclear spins of the host material (primarily $^{89}$Y$^{3+}$), and the electron spins of other nearby dopant $^{167}$Er$^{3+}$ ions. We first analyze the dependence of $T_2^{\rm hyp}$ on Er concentration ($n_{\rm Er}$) at zero field, clarifying the crucial tradeoff between memory time and readout efficiency. We then extend our search to non-zero magnetic fields to identify optimal ZEFOZ transitions, and evaluate their tolerance to experimental imperfections.

\section{Electron-nuclear spin coupled system}\label{sec2}

\subsection{Energy structure of $^{167}$Er$^{3+}$:Y$_2$SiO$_5$}\label{sec2-1}

The host crystal, Y$_2$SiO$_5$ (YSO), is known as a low magnetic noise material. This is because its constituent ions have either small magnetic moments or low natural abundances of magnetic isotopes (see Appendix~\ref{A1}). Consequently, magnetic fluctuations from the host crystal originate mainly from the $^{89}$Y nuclei, which have a nuclear spin $I=1/2$ and 100\% natural abundance.

In YSO, it is customary to use the optical frame, a coordinate system defined by the principal axes of polarization $D_1, D_2$, and $b$~\cite{petit2020demonstration}. The $^{167}$Er$^{3+}$ dopant, which is the only stable Er isotope with a non-zero nuclear spin ($I=7/2$), can substitute for Y$^{3+}$ ions at two distinct crystallographic sites, labeled site 1 and site 2. Both sites possess the lowest possible local symmetry, $C_1$, which leads to highly anisotropic magnetic properties.

\begin{figure}[th]
    \centering
    \includegraphics[width=1.0\columnwidth, clip]{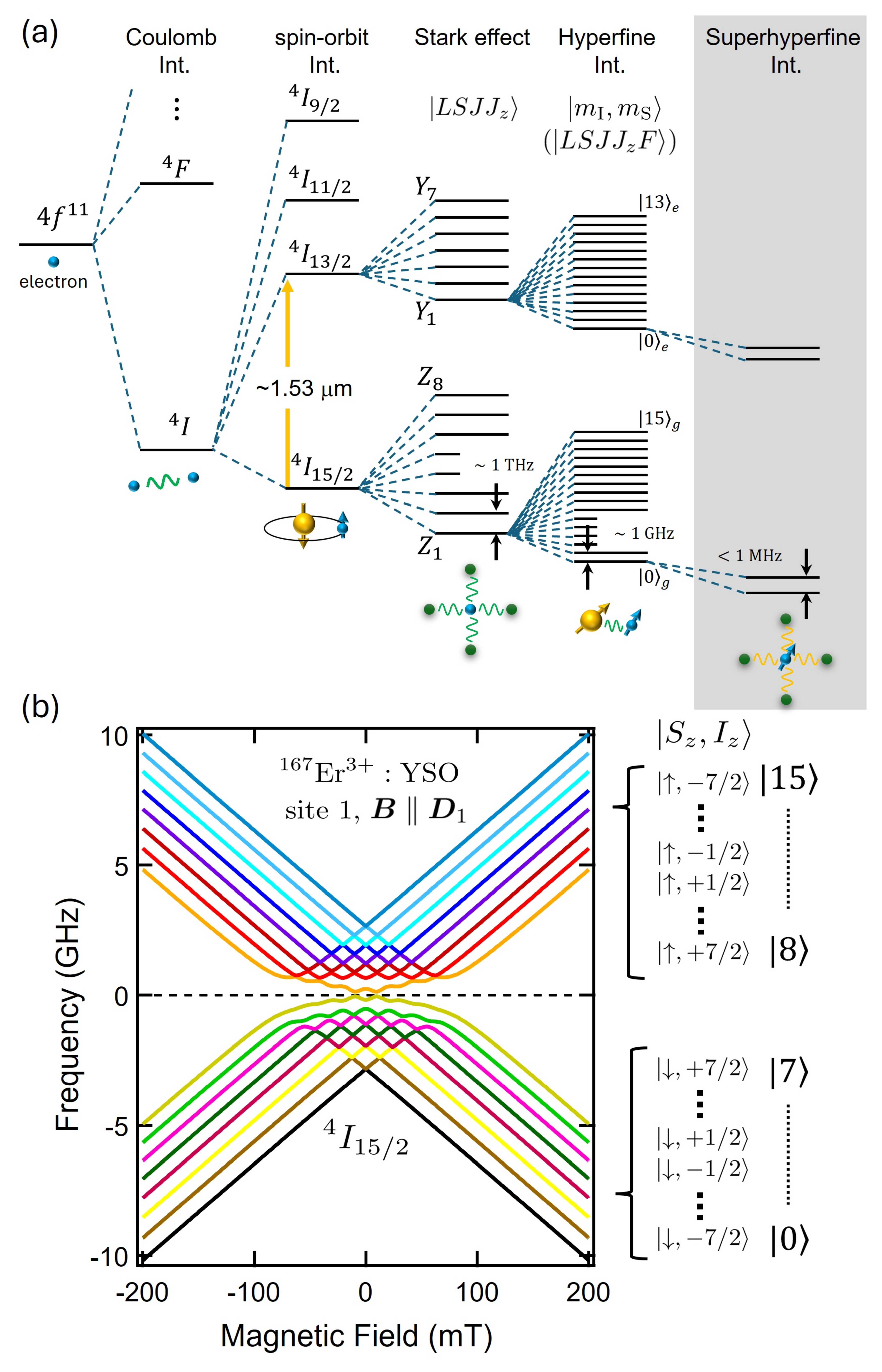}
\caption{(a) Schematic of the energy level structure of $^{167}$Er$^{3+}$ doped in YSO. The superhyperfine interaction, depicted in the gray shaded area, is neglected in the effective spin Hamiltonian of Eq.~\eqref{eqSpinH}. Its effect is not considered in the search for ZEFOZ points, as its magnitude is negligible compared to the hyperfine interaction. (b) The energy level structure of the ground states $^4 I_{15/2}\ (Z_1)$ of $^{167}$Er$^{3+}$:YSO at site 1. The external magnetic field $\bm B$ is applied along the $D_1$ axis. Each state can be expressed by $\ket{S_z, I_z}$ where $S_z=+1/2 \ (-1/2)$ is represented by $\uparrow \ (\downarrow)$ and is denoted from the lowest to highest as $\ket{i}$ ($i=0, 1, 2,\cdots, 15$) for simplicity.}
\label{Fig1}
\end{figure}

Each crystallographic site (1 and 2) is further composed of four magnetically distinct subsites for the Er$^{3+}$ ion, which arise from the crystal's $C_2$ rotation and inversion symmetries~\cite{sun2008magnetic}. Two of these subsites are related by inversion and are thus always magnetically equivalent, regardless of the magnetic field orientation. The other two are related by the $C_2$ rotation about the $b$-axis and become magnetically inequivalent when $\bm B$ is applied in a general direction. All four subsites become equivalent only when $\bm B$ is aligned with a high-symmetry direction: either parallel to the $b$-axis or lying within the $D_1-D_2$ plane. This subsite structure is essential for interpreting experimental spectra, because the number of observed transitions depends on the field orientation.

The complete energy structure of an Er$^{3+}$ ion in the crystal is described by a full Hamiltonian, which can be conceptually written as ${\cal H} = {\cal H}_{\rm FI} + {\cal H}_{\rm CF} + {\cal H}_{\rm spin}$. As illustrated in Fig.~\ref{Fig1}(a), the largest energy scales are set by the free-ion Hamiltonian ${\cal H}_{\rm FI}$ (Coulomb and spin-orbit interactions, $\sim$THz scale) and the crystal field Hamiltonian ${\cal H}_{\rm CF}$ (Stark effect, $\sim$GHz-THz scale)~\cite{Macfarlane1987book,liu2006spectroscopic}. The smaller magnetic interactions ${\cal H}_{\rm spin}$ (hyperfine, Zeeman and nuclear quadrupole terms, $\sim$MHz-GHz scale) determine the fine structure.

At the cryogenic temperatures relevant to this study, thermal energy is much smaller than the splitting between Stark levels. Consequently, only the lowest-energy Kramers doublet of the ground-state multiplet, $^4 I_{15/2}\ (Z_1)$, is populated. This allows us to simplify the problem by constructing an \textit{effective spin Hamiltonian}, ${\cal H}_{\rm spin}^{\rm eff}$, which operates exclusively within the Hilbert space of this ground-state doublet, treating it as an effective spin $S=1/2$ system. This model accurately describes the smaller magnetic interactions that lift the remaining degeneracy of the ground state. This effective spin Hamiltonian is given by:

\begin{align}
\begin{split}
{\cal H}_{\rm spin}^{\rm eff}&= \bm I \cdot \bm A \cdot \bm S + \bm I \cdot \bm Q \cdot \bm I \\
&+ \mu_{\rm B} \bm B \cdot {\rm {\mathbf g}}_{\rm e} \cdot \bm S
-  \mu_{\rm N} {\rm g}_{\rm n} \bm B \cdot \bm I. \label{eqSpinH}
\end{split}
\end{align}
Here, $\bm A$ is the HFI matrix, $\bm Q$ is the NQI matrix~\cite{comment1}, $\bm B$ is the applied magnetic field, ${\rm {\mathbf g}}_{\rm e}$ is the g-tensor of the EZI, $\mu_{\rm B}$ and $\mu_{\rm N}$ are the Bohr and nuclear magnetons, and ${\rm g}_{\rm n}$ is the nuclear g-factor. See Refs.~\onlinecite{mcauslan2012reducing,Macfarlane1987book} for a detailed discussion.

Figure~\ref{Fig1}(b) shows the energy structure of the $^4 I_{15/2}\ (Z_1)$ ground state for site 1, calculated using ${\cal H}_{\rm spin}^{\rm eff}$ with parameters from Ref.~\onlinecite{wang2023hyperfine}. These 16 coupled states are expressed in the basis $\ket{S_z, I_z}$, where $S_z$ and $I_z$ are the quantum numbers representing the projections of the effective electron spin $\bm S$ and nuclear spin $\bm I$, respectively, onto the quantization axis defined by the direction of the applied magnetic field $\bm B$. Even at zero magnetic field, the HFI and NQI induce strong state mixing, as indicated by the avoided level crossings. This state mixing suppresses the linear (first-order) Zeeman effect near zero field, causing the energy levels to instead exhibit a quadratic dependence on the magnetic field---a phenomenon known as the second-order Zeeman effect~\cite{analysis2005noel, mcauslan2012reducing}. As the magnetic field increases, the EZI becomes dominant, and the energy of each state shifts linearly.

The key parameters determining this fine energy structure are the $\bm A$, $\bm Q$, and ${\rm {\mathbf g}}_{\rm e}$ matrices. The high anisotropy of these tensors is a direct consequence of the low $C_1$ local symmetry at the Er substitution sites. This $C_1$ symmetry, the lowest possible, imposes no constraints on the relative orientation of the principal axes of these tensors; they are not required to align with each other or with any specific crystallographic axis~\cite{sun2008magnetic}. As will be shown later [Figs.~\ref{Fig6}(c) and~\ref{Fig7}(c)], this anisotropy is responsible for the existence of ZEFOZ points even at zero magnetic field. While several reports on these tensors exist with minor discrepancies ($\sim$0.1 GHz)~\cite{chen2018hyperfine,rakonjac2020long,jobbitt2021prediction,wang2023hyperfine}, for all calculations in this study, we use the set most recently refined by S-J. Wang \textit{et al.}~\cite{wang2023hyperfine}, as listed in Appendix~\ref{AQgMatrices}.

\subsection{Magnetic fluctuations in $^{167}$Er$^{3+}$:Y$_2$SiO$_5$}\label{sec2-2}

The spin coherence time $T_2^{\rm hyp}$ is fundamentally limited by the magnetic coupling of the target Er$^{3+}$ electron spin to the fluctuating magnetic fields produced by surrounding nuclear and electron spins~\cite{SpinPhysicsinSC,kaji2012direct}. To model this decoherence, we must first quantify the magnitude of these magnetic fluctuations, $\Delta \bm B$. They arise from two primary sources: the host crystal ions ($\Delta \bm B_{\rm host}$) and other dopant Er ions ($\Delta \bm B_{\rm Er}$). We evaluate the statistical distribution of these fluctuations using a Monte-Carlo simulation of the dipole-dipole interactions.

The total spin coherence time $T_2^{\rm hyp}$ is given by~\cite{MZhongthesis,Rancicgthesis}
	\begin{equation}
	\frac{1} {{\pi} T_2^{\rm hyp}} = \frac{1} {2{\pi}T_1^{\rm hyp}} + \gamma_\phi, \label{eq3}
	\end{equation}
where $T_1^{\rm hyp}$ is the population relaxation time and $\gamma_\phi$ is the pure dephasing rate. The latter is composed of three main contributions~\cite{equall1995homogeneous,MZhongthesis}:
	\begin{equation}
	\gamma_\phi=\gamma_{\rm Er-phonon}+\gamma_{\rm Er-host}+\gamma_{\rm Er-Er}. \label{eq4}
	\end{equation}
The phonon-induced term, $\gamma_{\rm Er-phonon}$, is strongly suppressed at the cryogenic temperatures ($<$ 2~K) considered in this study~\cite{rancic2018coherence}. We therefore focus on the two magnetic contributions, $\gamma_{\rm Er-host}$ and $\gamma_{\rm Er-Er}$, which are driven by the magnetic fluctuations $|\Delta \bm B_{\rm Y}|$ and $|\Delta \bm B_{\rm Er}|$, respectively.

For the Er-host interaction, the magnetic fluctuation from $^{89}$Y nuclei ($\sim 4.45 \ \mu$T) is over an order of magnitude larger than the contributions from $^{29}$Si and $^{17}$O, which are negligible ($\sim 0.1 \ \mu$T) due to their small natural abundances. We therefore consider only the effect of Y$^{3+}$. For the Er-Er interaction, we similarly consider only fluctuations from surrounding Er electron spins, because the nuclear magnetic moment of Er is much smaller than its electron magnetic moment.

The magnetic fluctuation $\Delta \bm B_{\rm A}$ (where A = Y or Er) acting on the target Er electron is generated by the dipole-dipole interaction with N surrounding spins:
  \begin{align}
    \Delta \bm B_{\rm A}&=\frac{\mu_0}{4 \pi} \sum_{i{=}1}^{N} \left[ \frac{3(\bm \mu_{{\rm A}_i} \cdot \bm r_i )\bm r_i}{ |\bm r_i|^5}-\frac{\bm \mu_{{\rm A}_i}}{| \bm r_i|^3}\right].
    \label{eq6}
  \end{align}
Here, $\bm r_i$ is the position vector from the target Er$^{3+}$ to the $i$-th spin, and $\bm \mu_{{\rm A}_i}$ is the magnetic moment of that spin. For Y nuclei, $\bm \mu_{{\rm Y}_i}$ is calculated assuming a random orientation, with its magnitude given by the gyromagnetic ratio in Appendix~\ref{A1}. For other Er ions, $\bm \mu_{{\rm Er}_i}$ is given by $\mu_{\rm B} {\mathbf  g}_{\rm e}\cdot \bm S$, where the electron spin momentum $\bm S$ is assumed to be randomly oriented.

\begin{figure}[t]
  \centering
    \includegraphics[width=1.0\columnwidth,clip]{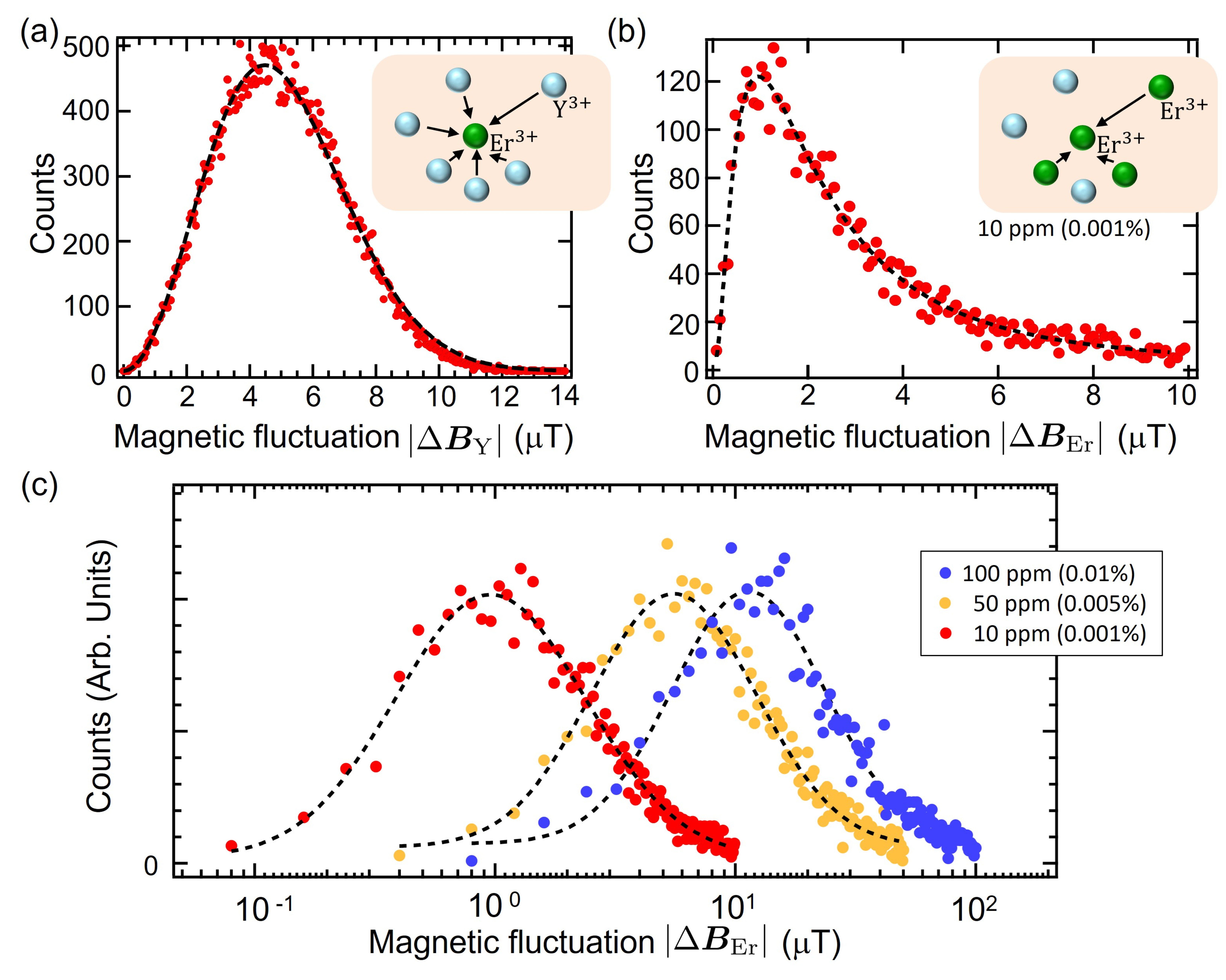}
\caption{Histogram of magnetic fluctuation by (a) Y ions $|\Delta \bm B_{\rm Y}|$ and (b) doped Er ions ($n_{\rm Er}= 10$ ppm) $|\Delta \bm B_{\rm Er}|$. The black dashed lines are fitting by probability density function. Insets show the images of magnetic fluctuations acting on the target Er$^{3+}$ center. (c) Dependence of $|\Delta \bm B_{\rm Er}|$ on $n_{\rm Er}$. The black dashed lines are the fitting curves. The distributions are normalized at their respective peak values. }
\label{Fig2}
\end{figure}

Figures~\ref{Fig2}(a) and~\ref{Fig2}(b) show the computed histograms of the magnitudes $|\Delta \bm B_{\rm Y}|$ and $|\Delta \bm B_{\rm Er}|$ (for a representative $n_{\rm Er}=10$~ppm) from 6000 Monte-Carlo iterations. The distributions show peaks at 4.45~$\mu$T and $\sim$1~$\mu$T, respectively. The histograms are well-fitted by a Maxwell-Boltzmann distribution (black dashed lines), which is expected for the magnitude of a 3D vector whose components are Gaussian-distributed, as is the case for the sum of many randomly oriented spins. For all subsequent analysis, we take the most probable value (the peak of the histogram) as the effective magnitude of the magnetic fluctuation, $|\Delta \bm B|$. The full width at half maximum (FWHM) of the Y fluctuation derived from this peak value is comparable to previously reported values~\cite{MZhongthesis}.

Figure~\ref{Fig2}(c) shows the dependence of $|\Delta \bm B_{\rm Er}|$ on the Er concentration, $n_{\rm Er}$. As $n_{\rm Er}$ increases, the average inter-Er distance decreases, leading to a corresponding increase in $|\Delta \bm B_{\rm Er}|$. The specific values obtained from these distributions are used in the following sections to calculate the spin coherence times at ZEFOZ transitions.

\section{Exploration of optimal ZEFOZ transitions in $^{167}$E\MakeLowercase{r}$^{3+}$:YSO}\label{sec3}
\subsection{Er concentration dependence of spin coherence time at zero magnetic field}\label{sec3-1}

Here, we investigate the dependence of the spin coherence time $T_2^{\rm hyp}$ on the $^{167}$Er$^{3+}$ concentration $n_{\rm Er}$ under zero magnetic field. Even at ${\bm B}={\bm 0}$, ZEFOZ transitions exist where the first-order sensitivity to magnetic fields, $\bm S_1$, vanishes~\cite{ortu2018simultaneous,rakonjac2020long}. At these points, $T_2^{\rm hyp}$ is limited by the second-order sensitivity (curvature) $\bm S_2$ and the magnitude of the magnetic fluctuations $\Delta \bm B$, as modeled by Eq.~\eqref{eq8}. The details of the calculation method for $\bm S_1$ and $\bm S_2$ are provided in Appendix~\ref{A3}.

\begin{equation}
\frac{1}{\pi T_2^{\rm hyp}}=\bm S_1 \cdot (2\Delta \bm B) + (2\Delta \bm B) \cdot \bm S_2 \cdot (2\Delta \bm B).
\label{eq8}
\end{equation}

\begin{figure}[tb]
  \centering
    \includegraphics[width=1.0\columnwidth, clip]{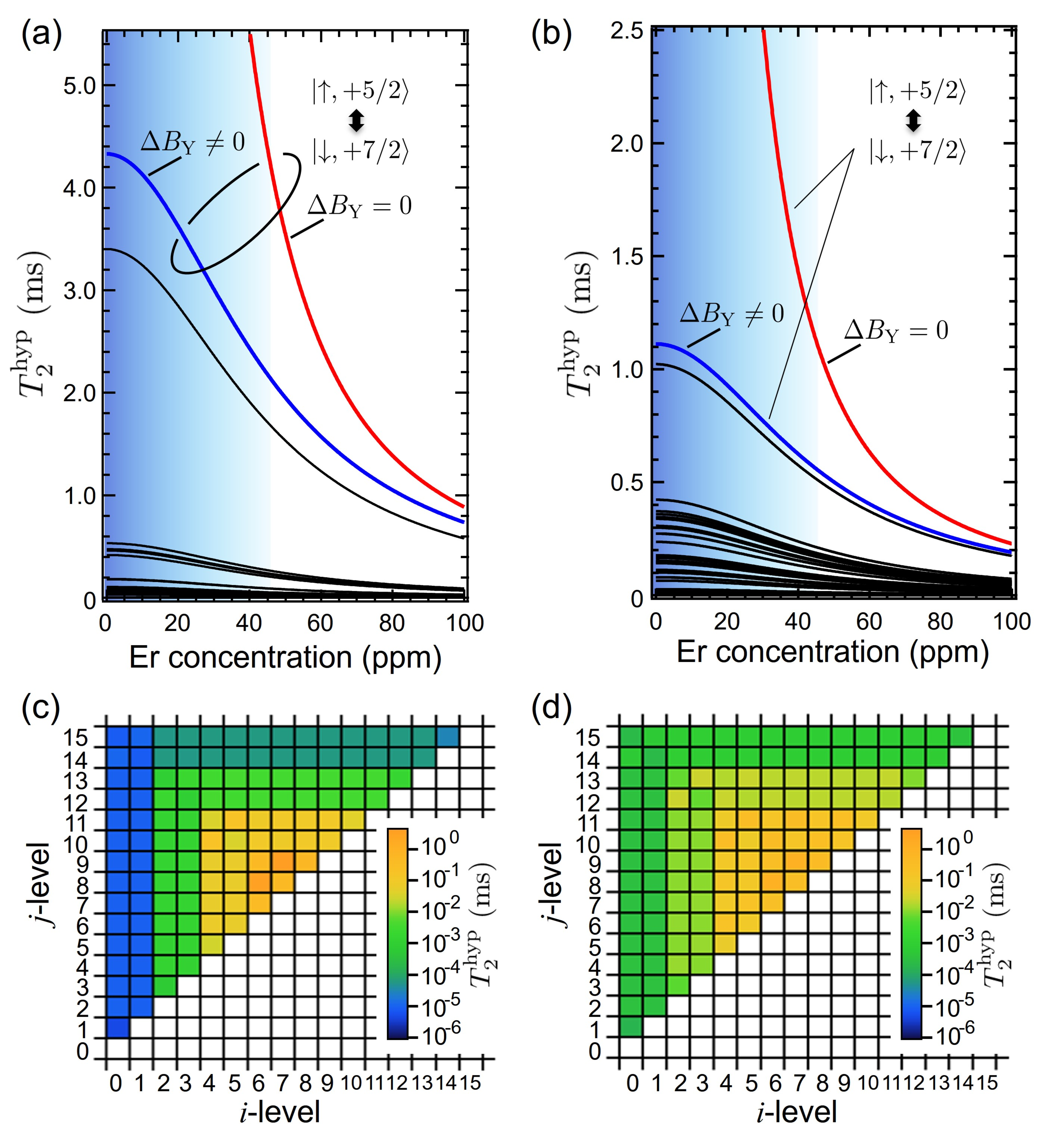}
\caption{Dependence of $T_2^{\rm hyp}$ on $n_{\rm Er}$ at 0 T for (a) site 1 and (b) site 2. The red and blue lines indicate the dependence for the ZEFOZ transition $\ket{7} \rightleftarrows \ket{9} \ (\ket{\downarrow,+7/2} \rightleftarrows \ket{\uparrow,+5/2})$ for site 1 and site 2 affected by only $\Delta \bm B_{\rm Er}$ and by $\Delta \bm B_{\rm Er}$ and $\Delta \bm B_{\rm Y}$, respectively. The black solid lines represent the dependence of $T_2^{\rm hyp}$ on $n_{\rm Er}$ for other ZEFOZ transitions affected by $\Delta \bm B_{\rm Er}$ and $\Delta \bm B_{\rm Y}$. (c) $T_2^{\rm hyp}$ mapping for the ZEFOZ transitions between the $i$th and $j$th HF levels affected by $\bm B_{\rm Er}$ and $\Delta \bm B_{\rm Y}$ for (c) site 1 and (d) site 2 with the same color scale at $n_{\rm Er}$=10 ppm.}
\label{Fig3}
\end{figure}

Figures~\ref{Fig3}(a) and~\ref{Fig3}(b) show the calculated $n_{\rm Er}$-dependence of $T_2^{\rm hyp}$ for the longest-lived zero-field ZEFOZ transition ($\ket{7} \rightleftarrows \ket{9}$) for each site. The physical reason for the observed saturation is clearly illustrated by comparing two theoretical scenarios. The red curve, which considers only the concentration-dependent Er-Er interaction ($|\Delta \bm B_{\rm Er}|$), shows that $T_2^{\rm hyp}$ would hypothetically continue to increase as $n_{\rm Er}$ decreases. The blue curve, however, represents the realistic situation by including the constant magnetic fluctuation from the host Y nuclei ($|\Delta \bm B_{\rm Y}|$). At high concentrations ($> \ \sim$100~ppm), the Er-Er interaction dominates and the curves nearly overlap. As the concentration decreases, the constant noise from Y nuclei becomes the dominant limiting factor, causing $T_2^{\rm hyp}$ to plateau. This saturation becomes evident for concentrations below approximately 10~ppm, indicating that further dilution of Er ions yields no significant improvement in coherence time under zero field. This finding highlights that the application of an external magnetic field is essential to achieve coherence times significantly beyond the few-millisecond zero-field limit.

Figures~\ref{Fig3}(c) and~\ref{Fig3}(d) map the $T_2^{\rm hyp}$ for all 120 zero-field ZEFOZ transitions at a representative concentration of $n_{\rm Er}=10$~ppm. The calculated $T_2^{\rm hyp}$ distributes over a wide range, from 8~ns to 4.12~ms for site 1 and 169~ns to 1.06~ms for site 2. A notable feature in these maps is the clustering of long-$T_2^{\rm hyp}$ transitions along the anti-diagonal ($j=15-i$). This is a direct consequence of the system's symmetry at zero magnetic field. As seen in Fig.~\ref{Fig1}(b), the 16 hyperfine levels form 8 pairs of states ($\ket{i}$ and $\ket{15-i}$) that are symmetrically arranged. For transitions between these symmetrically-paired states, the first-order Zeeman shifts of the two levels are often nearly identical, fulfilling the ZEFOZ condition (${dE_i}/{dB} \approx {dE_{15-i}}/{dB}$) and thus leading to enhanced coherence.

Selecting an optimal $n_{\rm Er}$ requires balancing the crucial tradeoff between spin coherence time, which favors lower concentrations, and readout efficiency, which is enhanced by the increased optical absorption at higher concentrations. While high efficiencies have been reported in 50-ppm crystals with protocols like 4-level photon echo~\cite{stuart2024progress} and Revival of Silenced Echo (ROSE)~\cite{minnegaliev2022implementation}, recent work shows that geometric optimization, such as extending sample length, can significantly improve efficiency even at 10~ppm~\cite{yasuiISNTT2024}. This suggests it is possible to maintain high efficiency in low-concentration samples, precisely where the longest spin coherence times are expected.

A key mechanism protecting the spin coherence is the frozen core effect~\cite{mims1968phase}, which suppresses host spin flip-flops. In $^{167}$Er:YSO, this effect is driven by the large, intrinsic magnetic moment of the Er \textit{electron} spin, making it potent even at zero magnetic field. This contrasts sharply with the benchmark system, $^{151}$Eu:YSO, where the frozen core is induced by the much weaker \textit{nuclear} magnetic moment. We define the frozen core radius as the distance at which the magnetic dipole-dipole interaction between a Y nucleus and a nearby Er electron spin becomes comparable to that between Y nuclear spins.
In other words, it is determined by the condition $|{\cal H}^{\rm Y-Er}_{\rm dd}| = |{\cal H}^{\rm Y-Y}_{\rm dd}|$. From Eq. (5), this condition can be equivalently expressed as $|\Delta \bm B_{\rm Er}| = |\Delta \bm B_{\rm Y}|$, where the magnitudes of the magnetic field fluctuations produced by the Er and Y ions are equal, which, from Fig.~\ref{Fig3}, corresponds to $n_{\rm Er} \approx 45.3$~ppm. Substituting this into Eq.~\eqref{eq9a} yields a radius of $r \approx 111.25$~{\AA}, encompassing approximately 108,000 Y$^{3+}$ ions. The validity of this remarkably large number of  Y$^{3+}$ ions is confirmed by a scaling argument: the Er electron magnetic moment ($\approx 7.35 \mu_B$) is over 1,000 times larger than the induced nuclear moment of Eu at $\sim$1~T~\cite{zhong2015optically}. Since the frozen core volume scales with the moment's strength, the number of included ions should be roughly three orders of magnitude larger, consistent with our calculation.

Finally, while our model predicts the properties of ideal ZEFOZ transitions, experimental results can be more complex. A counter-intuitive result reported by J. V. Rakonjac \textit{et al.}~\cite{rakonjac2020long} showed a shorter $T_2^{\rm hyp}$ for a ZEFOZ transition compared to a non-ZEFOZ transition at 0~T. Our calculations suggest that a small, uncompensated stray magnetic field of $\sim 1$~mT is a plausible explanation for this discrepancy (see Appendix~\ref{A4}).

\begin{figure}[t]
  \centering
    \includegraphics[width=0.9\columnwidth, clip]{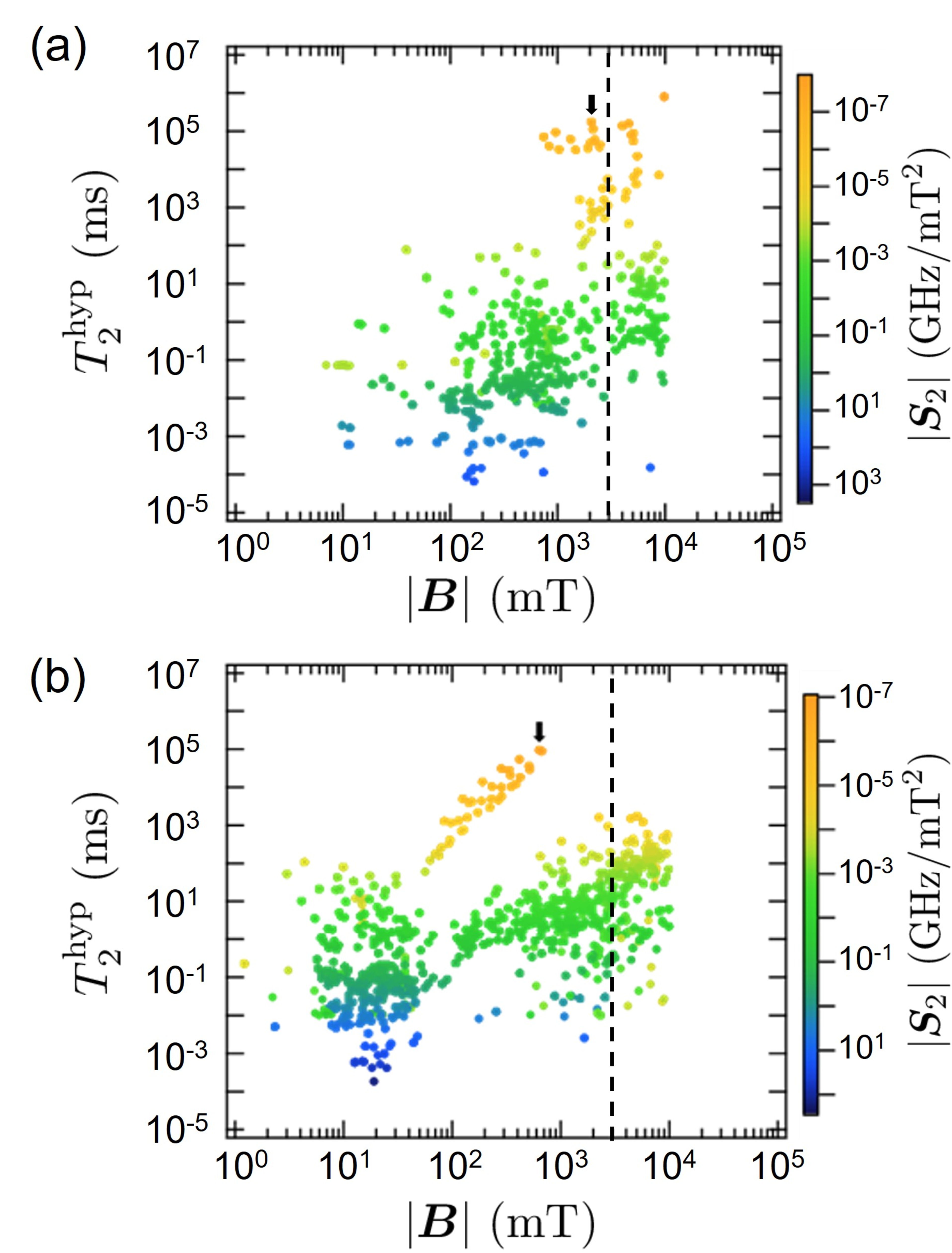}
\caption{$T_2^{\rm hyp}$ and $|\bm B|$ of the obtained ZEFOZ points for (a) site 1 and (b) site 2. The color tables are scaled by $|\bm S_2|$. The transition with the longest $T_2^{\rm hyp}$ has the minimal $\bm S_2$ actually according to Eq.~\eqref{eq8}. The ZEFOZ transition with the longest $T_2^{\rm hyp}$ under our condition $|\bm B|\le 3$ T corresponds to $\ket{10}-\ket{11}$ $[\ket{\uparrow,+1/2} \rightleftarrows \ket{\uparrow,+3/2}]$ ($\ket{14}-\ket{15}$ $[\ket{\uparrow,-7/2} \rightleftarrows \ket{\uparrow,-5/2}]$) for site 1 (site 2) indicated by the arrow. The dashed vertical line indicates $|\bm B|=3$ T.}
\label{Fig4}
\end{figure}

\subsection{Spin coherence time under non-zero magnetic field}\label{sec3-2}

The previous section found that zero-field ZEFOZ transitions limit $T_2^{\rm hyp}$ to the range of a few milliseconds at most for both site 1 and site 2. To achieve significantly longer coherence times, the application of an external magnetic field is essential. This section investigates the $T_2^{\rm hyp}$ of ZEFOZ transitions in the $^4I_{15/2} \ (Z_1)$ state of $^{167}$Er$^{3+}$:YSO (with $n_{\rm Er}=10$ ppm) under non-zero magnetic fields. The ZEFOZ points were identified using a systematic numerical search based on the Newton method, as detailed in Appendix~\ref{A3}.

Figure~\ref{Fig4} plots the calculated $T_2^{\rm hyp}$ for $n_{\rm Er}=10$ ppm as a function of the applied field magnitude $|\bm B|$. Since the application of a strong field ($> 3$ T) would eliminate the population of the spin-up electron levels, our search was restricted to $|\bm B| \le 3$ T. The optimal transitions identified (see Appendix~\ref{A5} for a detailed list) are $\ket{10}\rightleftharpoons\ket{11}$ for site 1 and $\ket{14}\rightleftharpoons\ket{15}$ for site 2, indicated by arrows in Fig.~\ref{Fig4}. These points yield maximum predicted $T_2^{\rm hyp}$ of 173.9 s and 95.3 s, respectively. This represents an improvement of more than four orders of magnitude compared to the zero-field limit.

\begin{figure}[t]
  \centering
    \includegraphics[width=1.0\columnwidth, clip]{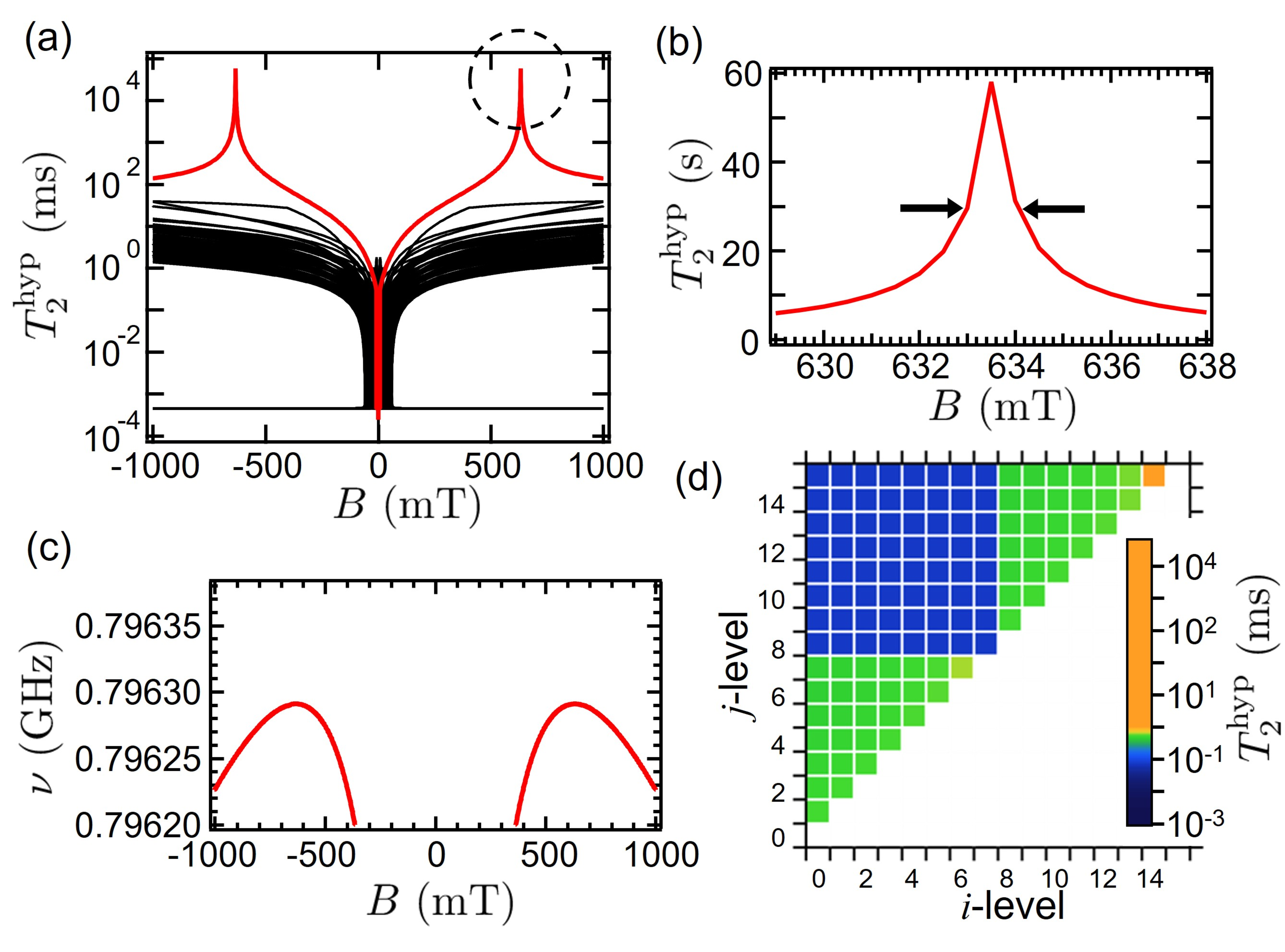}
\caption{(a) The magnetic field response of the ZEFOZ point with the longest $T_2^{\rm hyp}$ at site 2 is indicated by the red line, and the response of other ZEFOZ points are shown by the black lines. The magnetic field direction for the ZEFOZ point with the longest $T_2^{\rm hyp}$ is $(\theta, \phi)=$( $\pm$37.5383$^{\circ}$, -10.9417$^{\circ}$).  (b) Enlarged view of the dashed circle in (a) around +0.634 T. The FWHM of the peak is $\sim$1 mT. (c) Magnetic field magnitude $B$ versus transition frequency $\nu$ for the ZEFOZ point with the longest $T_2^{\rm hyp}$ (red line). (d) $T_2^{\rm hyp}$ mapping for the transition between the $i$th and $j$th HF levels at $B$=+633.52 mT.
}
\label{Fig5}
\end{figure}

As a general trend observed in Fig.~\ref{Fig4}, longer $T_2^{\rm hyp}$ values are predominantly found at higher magnetic fields. This can be understood by considering two distinct regimes. In the weak-field regime (e.g., $|\bm B| < \sim$200 mT), the Zeeman interaction is comparable in strength to the HFI, leading to strong state mixing and numerous avoided level crossings [see Fig.~\ref{Fig1}(b)]. This results in a large energy level curvature ($|\bm S_2|$) and consequently, shorter $T_2^{\rm hyp}$. In contrast, in the strong-field regime (e.g., $|\bm B| > 1$ T), the electronic Zeeman interaction dominates. This causes the energy levels to decouple into two manifolds with a more linear field dependence, significantly reducing the curvature $|\bm S_2|$ and allowing for much longer $T_2^{\rm hyp}$.

This trend is particularly evident for site 1, where most ZEFOZ points with long $T_2^{\rm hyp}$ are found in the strong-field regime ($|\bm B| > 1$ T). However, operating at such high fields introduces a potential complication: mixing with higher-lying Stark levels. As the Zeeman shift becomes comparable to the crystal field splitting, crossings between the ground (Z$_1$) and first excited (Z$_2$) Stark levels can occur~\cite{jobbitt2021prediction}, invalidating the 16-level effective spin Hamiltonian model. Since the optimal ZEFOZ point for site 2 is found at a more moderate field ($\sim$0.6 T), it is not subject to this issue. This suggests that site 2 may offer a more straightforward and reliable platform for experimental implementation.

As site 2 is a promising candidate, we analyze one of its optimal ZEFOZ point ($\ket{14}-\ket{15}$) in detail. The main parameters for this transition are as follows:
\begin{itemize}
	\item $\bm B_{(D_1, D_2, b)}=$($\mp$378.9,  $\pm$73.2,  $\mp$502.3)  mT
	\item $\bm B_{(B, \theta, \phi)}=$(633.52 mT, $\mp$37.5383$^{\circ}$, -10.9417$^{\circ}$)
	\item $|\bm S_1|=5.46379 \times 10^{-13}$ GHz/mT
	\item $|\bm S_2|_{\rm max}=9.90 \times 10^{-8}$ GHz/mT$^2$
\end{itemize}
Here, the upper (lower) sign of $\bm B_{(D_1, D_2, b)}$ and $\theta$ corresponds to orientation 1 (orientation 2) for site 2. This point clearly satisfies the ZEFOZ condition. As shown in Figs.~\ref{Fig5}(a) and~\ref{Fig5}(c), the transition frequency $\nu$ exhibits a clear turning point at $|\bm B|\approx 633.5$ mT, which corresponds to the sharp peak in $T_2^{\rm hyp}$. This optimal point is also practical: Fig.~\ref{Fig5}(b) shows that the coherence time remains above half its peak value even with a $\sim$1 mT error in the field magnitude.

Figure~\ref{Fig5}(d) shows the $T_2^{\rm hyp}$ distribution across all level pairs at this optimal field. In the strong-field regime, the 16 levels decouple into two manifolds: spin-down ($\ket{0} - \ket{7}$) and spin-up ($\ket{8} - \ket{15}$) [see Fig.~\ref{Fig1}(b)]. The physical origin of the observed pattern is that transitions within the same manifold (e.g., up-up) involve states with strongly correlated Zeeman shifts, making it probable to find conditions where they cancel ($dE_i/dB \approx dE_j/dB$), thus satisfying the ZEFOZ condition. Conversely, transitions between the two manifolds (down-up) have large, opposite shifts, making them inherently sensitive to magnetic noise and extremely unlikely to be ZEFOZ points.

The overall coherence $T_2^{\rm hyp}$ is ultimately limited not only by dephasing ($\gamma_\phi$) but also by the population relaxation time $T_1^{\rm hyp}$, as shown in Eq.~\eqref{eq3}. This relaxation is governed by spin-phonon interactions. To achieve the predicted multi-second $T_2^{\rm hyp}$, $T_1^{\rm hyp}$ must be extended by cooling the crystal. Based on Ref.~\onlinecite{Rancicgthesis}, a temperature of approximately 0.14 K is required to suppress the direct phonon process for the relevant GHz-scale electron spin splittings (94.6 GHz for site 1, 68.1 GHz for site 2). This required temperature exceeds the $T < 100$ mK regime where the phonon bottleneck effect is known to dominate relaxation in Er:YSO~\cite{budoyo2018phonon}. Therefore, this bottleneck effect is not expected to be a primary obstacle, and we conclude that the predicted coherence times are experimentally feasible with sufficient cooling.

\begin{figure}[t]
  \centering
    \includegraphics[width=1.0\columnwidth, clip]{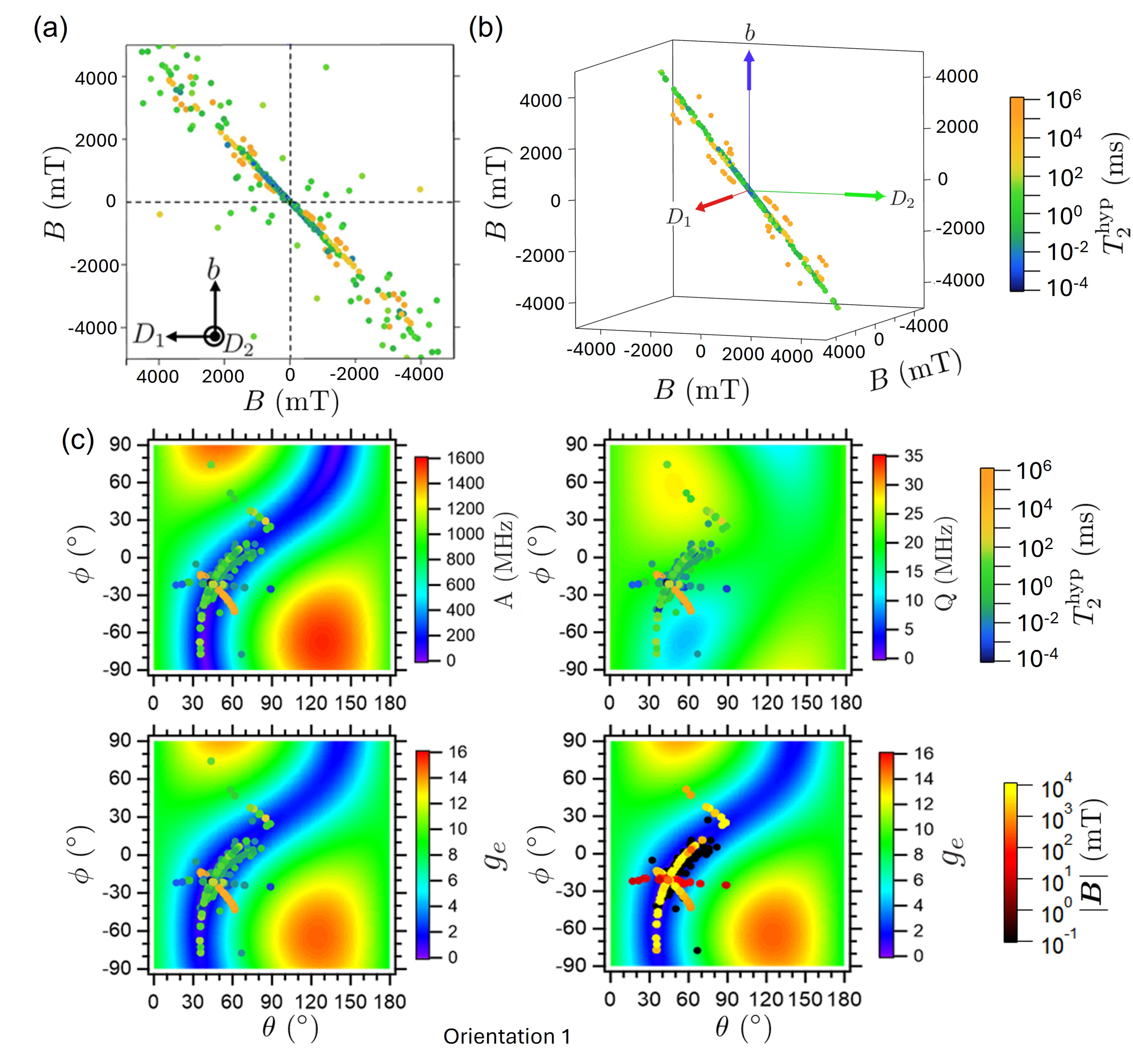}
\caption{(a), (b) Spatial distribution of ZEFOZ points for site 1 in the magnetic field frame $B_{(D_1, D_2, b)}$. Color distinguishes long ($T_2^{\rm hyp} > 10^3$ ms, orange) from short ($T_2^{\rm hyp} < 10^3$ ms, green/blue) coherence times. (c) Anisotropy patterns of the $\bm A$ (top left), $\bm Q$ (top right), and ${\rm {\mathbf g}}_{\rm e}$ (bottom left) tensors for orientation 1 ($\theta > 0$). Color represents the tensor component's magnitude along the $(\theta, \phi)$ direction (MHz for $\bm A$, $\bm Q$; dimensionless for ${\rm {\mathbf g}}_{\rm e}$). Patterns for other orientations can be inferred from crystal symmetry. ZEFOZ points are superimposed using the same color scheme as in (a) and (b). The bottom right panel instead colors the ZEFOZ points by the applied field magnitude, $|\bm B|$.
}
\label{Fig6}
\end{figure}

Plotting the spatial coordinates of the identified ZEFOZ points reveals distinct geometric patterns. For site 1, the points show a strong linear alignment [Figs.~\ref{Fig6}(a) and~\ref{Fig6}(b)], a phenomenon previously reported~\cite{wang2023hyperfine}. To understand this, we examined the angular dependence of the spin Hamiltonian tensors [Fig.~\ref{Fig6}(c)]. The patterns show that the principal axes of the electronic interaction tensors ($\bm A$ and ${\rm {\mathbf g}}_{\rm e}$) are closely aligned, while the nuclear quadrupole tensor ($\bm Q$) is oriented differently. This misalignment is a direct consequence of the low $C_1$ local symmetry, which permits maximum anisotropy~\cite{guillot2006hyperfine}. When the ZEFOZ points are overlaid on this data, those with long $T_2^{\rm hyp}$ cluster at specific angles ($\theta \sim 50^{\circ}, \phi \sim -30^{\circ}$), while shorter $T_2^{\rm hyp}$ points trace the valleys where the electronic interaction tensors are weakest.

\begin{figure}[t]
  \centering
    \includegraphics[width=1.0\columnwidth, clip]{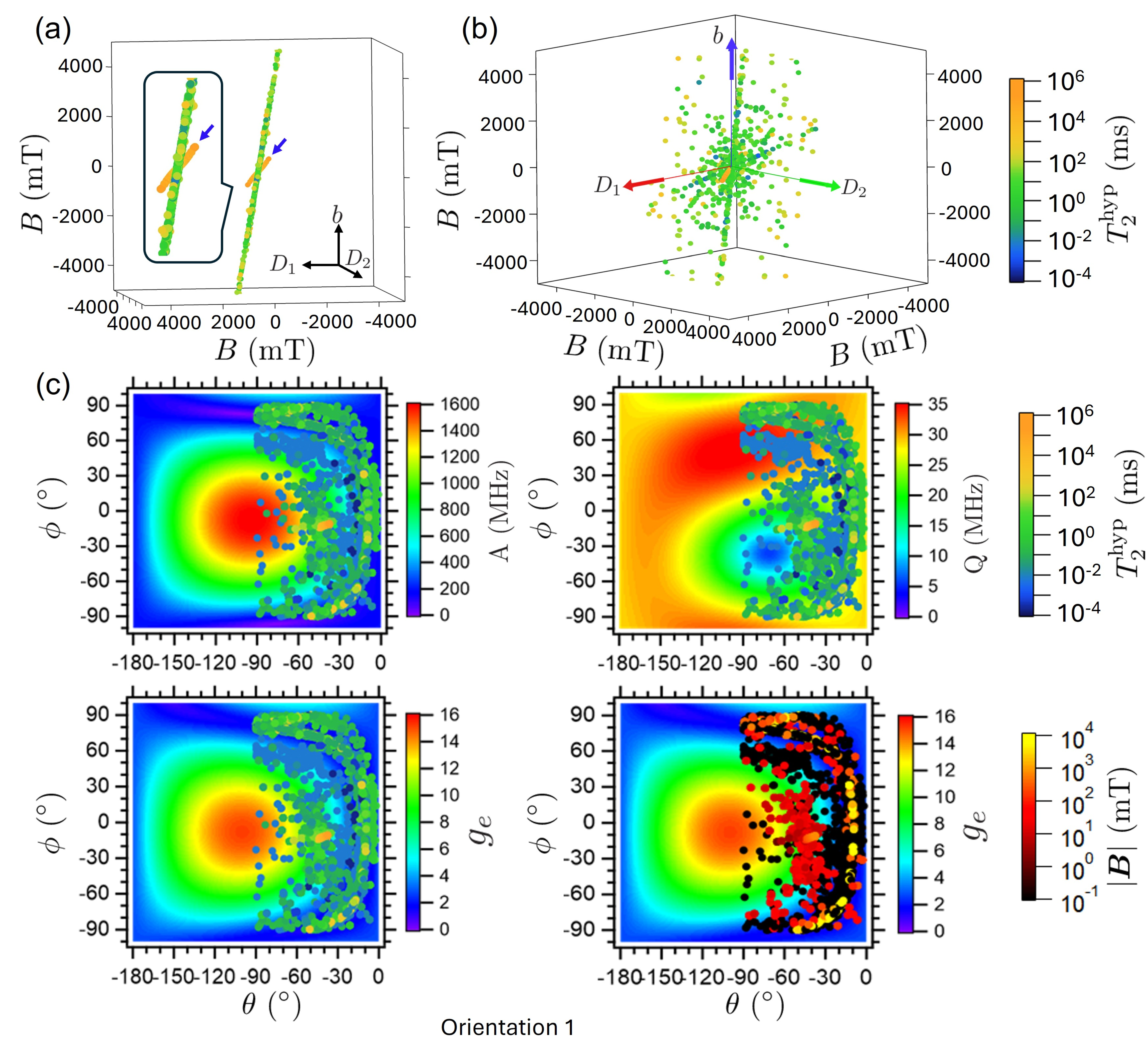}
\caption{(a), (b) Spatial distribution of ZEFOZ points for site 2 in the magnetic field frame $B_{(D_1, D_2, b)}$. Color distinguishes long ($T_2^{\rm hyp} > 10^3$ ms, orange) from short ($T_2^{\rm hyp} < 10^3$ ms, green/blue) coherence times. While most points lie within a plane, a line of long-coherence ZEFOZ points (blue arrow) deviates in the weak-field region. (c) Anisotropy patterns of the $\bm A$ (top left), $\bm Q$ (top right), and ${\rm {\mathbf g}}_{\rm e}$ (bottom left) tensors for orientation 1 ($\theta < 0$). Color represents the tensor component's magnitude along the $(\theta, \phi)$ direction (MHz for $\bm A$, $\bm Q$; dimensionless for ${\rm {\mathbf g}}_{\rm e}$). Patterns for other orientations can be inferred from crystal symmetry. ZEFOZ points are superimposed using the same color scheme as in (a) and (b). The linear feature from (a) corresponds to the cluster at $(\theta, \phi)\sim (-40^{\circ}, -10^{\circ})$. The bottom right panel instead colors the ZEFOZ points by the applied field magnitude, $|\bm B|$.
}
\label{Fig7}
\end{figure}

Additionally, the ZEFOZ points for site 2 were investigated [Figs.~\ref{Fig7}(a) and (b)]. These points are approximately distributed within a single plane with the normal vector ${\bm n}$ = (0.702, 0.532, -0.471). However, a distinct line of high-coherence points (indicated by the blue arrow in Fig.~\ref{Fig7}(a)) deviates from this plane in the weak-field region, corresponding to the cluster around $(\theta, \phi) \sim (-40^{\circ}, -10^{\circ})$ in Fig.~\ref{Fig7}(c).

The physical origin of these distinct (linear vs. planar) distributions lies in the competition between anisotropic terms in the effective spin Hamiltonian, where the dominant term changes with field strength as shown in the bottom right panel of Fig.~\ref{Fig7}(c).
\begin{enumerate}
	\item \textbf{Strong-field regime (B $>$ $\sim$1 T, yellow dots)}: The EZI dominates, so ZEFOZ points cluster near the minima of the ${\rm {\mathbf g}}_{\rm e}$ tensor.
	\item \textbf{Intermediate-field regime ($\sim$0.01 T $<$ B $<$ 1 T, orange and red dots)}: The NQI ($\bm Q$ tensor) influence becomes comparable, and ZEFOZ points shift to angles that represent a compromise between the competing tensors.
	\item \textbf{Near-zero-field regime (B $< \sim$0.01 T, black dots)}: The field direction is ill-defined, and the distribution is dictated by the complex interplay of the zero-field terms ($\bm A$ and $\bm Q$).
\end{enumerate}


\begin{figure}[t]
  \centering
    \includegraphics[width=1.0\columnwidth, clip]{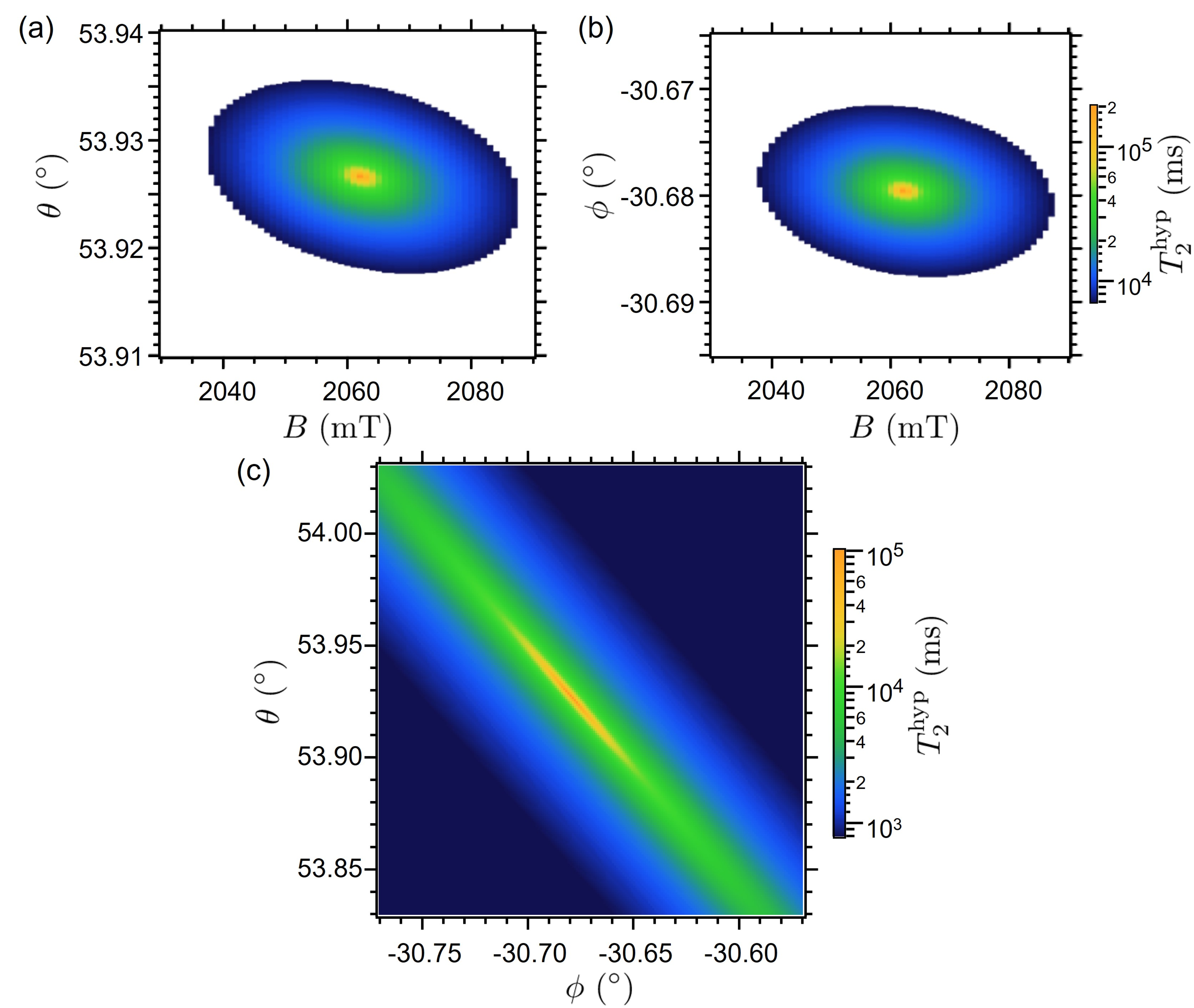}
\caption{$T_2^{\rm hyp}$ dependence on magnetic field parameters for the optimal ZEFOZ point of site 1 ($T_2^{\rm hyp} \sim 173.9$ s). (a) $T_2^{\rm hyp}$ versus field strength $B$ and polar angle $\theta$. (b) $T_2^{\rm hyp}$ versus $B$ and azimuth angle $\phi$. (c) $T_2^{\rm hyp}$ versus only the applied angles $\theta$ and $\phi$ at a constant field strength ($B=2.062$ T).
}
\label{Fig8}
\end{figure}

\begin{figure}[t]
  \centering
    \includegraphics[width=1.0\columnwidth, clip]{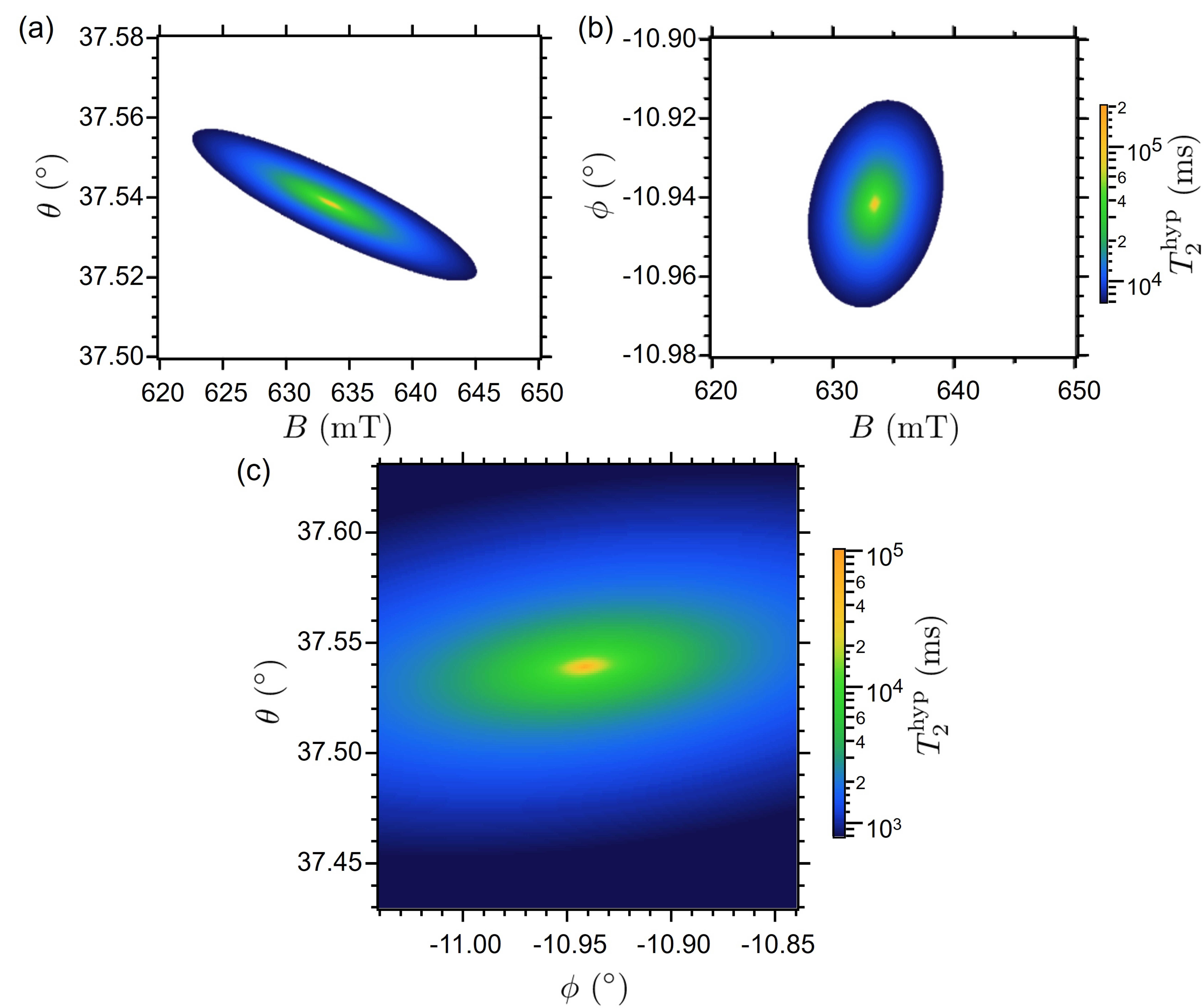}
\caption{$T_2^{\rm hyp}$ dependence on magnetic field parameters for the optimal ZEFOZ point of site 2 ($T_2^{\rm hyp} \sim 95.3$ s). (a) $T_2^{\rm hyp}$ versus field strength $B$ and polar angle $\theta$. (b) $T_2^{\rm hyp}$ versus $B$ and azimuth angle $\phi$. (c) $T_2^{\rm hyp}$ versus only the applied angles $\theta$ and $\phi$ at a constant field strength ($B \approx 634$ mT).
}
\label{Fig9}
\end{figure}

Finally, we assess the practical tolerance of these optimal points to errors in the applied magnetic field vector (Figs.~\ref{Fig8} and~\ref{Fig9}). This analysis reveals a critical trade-off. Site 1 offers the highest performance ($T_2^{\rm hyp} \approx 173.9$ s) but demands extreme precision. As shown in Figs.~\ref{Fig8}(a) and~\ref{Fig8}(b), an angular deviation of just 0.005$^{\circ}$ is sufficient to reduce $T_2^{\rm hyp}$ by one order of magnitude. This extreme sensitivity is further emphasized in Fig.~\ref{Fig8}(c), where the high-coherence region is confined to a narrow linear path; a deviation of only 0.05$^{\circ}$ perpendicular to this path causes the coherence time to drop by two orders of magnitude.

In stark contrast, site 2 presents a more pragmatic and robust target. While its maximum predicted $T_2^{\rm hyp}$ is shorter at $\sim$95.3 s, it offers far greater tolerance to field misalignment. As derived from Figs.~\ref{Fig9}(a)-(c), the angular deviation that reduces $T_2^{\rm hyp}$ by one order of magnitude is approximately 0.02$^{\circ}$. This tolerance is four times larger than the highly restrictive 0.005$^{\circ}$ limit of site 1. This superior angular robustness, visually confirmed by the large elliptical tolerance area in Fig.~\ref{Fig9}(c), significantly relaxes the experimental constraints. This makes site 2 a more feasible and promising candidate for the initial experimental demonstration of multi-second, telecom-compatible quantum memories.

\section{Conclusion}
In this study, we systematically explored ZEFOZ transitions in $^{167}$Er:YSO to identify optimal conditions for extending the spin coherence time ($T_2^{\rm hyp}$) for spin-wave storage. Our analysis of magnetic fluctuations confirmed that at zero field, $T_2^{\rm hyp}$ saturates for Er concentrations below 10 ppm due to constant noise from the host lattice, highlighting the necessity of applying an external magnetic field.

By extending our search to non-zero magnetic fields, we identified optimal ZEFOZ points where the predicted $T_2^{\rm hyp}$ is dramatically enhanced, exceeding 170 s for site 1 and 90 s for site 2—an improvement of over four orders of magnitude compared to the zero-field limit. We discovered that these optimal points form distinct geometric patterns—a line for site 1 and a plane for site 2—which we attribute to the high anisotropy of the spin Hamiltonian tensors arising from the low $C_1$ local symmetry.

Finally, our tolerance analysis highlights that realizing these ultra-long coherence times requires stringent angular control of the magnetic field, on the order of 0.005$^\circ$ for site 1. While experimentally demanding, this level of precision is not unprecedented and has been demonstrated in other rare-earth systems such as $^{151}$Eu$^{3+}$:YSO~\cite{zhong2015optically,MZhongthesis} and $^{171}$Yb$^{3+}$:YSO~\cite{ortu2018simultaneous} . This indicates that the optimal ZEFOZ points identified in this work are not merely theoretical ideals, but represent tangible targets for the experimental demonstration of ultra-long coherence times and spin-wave storage in a telecom-compatible quantum memory.

\begin{acknowledgments}
We thank Dr. Jevon Longdell of University of Otago for many valuable discussions and suggestions. We also thank Dr. X. Xu and Dr. H. Sanada of NTT Basic Research Laboratories for helpful discussion.
This work was supported by Japan Society for the Promotion of Science (JSPS) KAKENHI Grant Numbers JP23K17883, JP23K23263, and JP25K21708.
\end{acknowledgments}


\hrulefill

\begin{appendix}
\section{Isotopes of Er$^{3+}$:Y$_2$SiO$_5$} \label{A1}

As shown in Table~\ref{TBL1A}, Er, Si, and O ions have six, three, and three isotopes, respectively. Y ions have the only one isotope. 
Among these isotopes, $^{89}$Y, $^{167}$Er, $^{29}$Si, and $^{17}$O have non-zero nuclear spins.
$^{29}$Si and $^{17}$O have large nuclear magnetic moments $\bm \mu$ compared with $^{89}$Y, but small natural abundance. Thus, $^{167}$Er and $^{89}$Y are the main sources of the magnetic fluctuations. In this work, the magnetic fluctuation due to Er nuclei is neglected because the nuclear magnetic moment of Er is much smaller than the electron magnetic moment.

\begin{table}[h]
 \caption{Isotopes of Er$^{3+}$:Y$_2$SiO$_5$. $I$: nuclear spin quantum number, $\mu_{\rm N}$: nuclear magneton, $\gamma$: nuclear gyromagnetic ratio,  ${\rm g}_{\rm n}$: nuclear g-factor. The nuclear magnetic moment is given by $\bm \mu={\rm g}_{\rm n} \mu_{\rm N} \bm I=\gamma \hbar \bm I$, where $\hbar$ is the reduced Plank constant~\cite{NuclearElement}.}
 \begin{center}
  \begin{tabular}{|c|c|c|c|c|c|}
    \hline
    Element   & Abundance&  $I$  & $-|\bm \mu| / \mu_{\rm N}$    & $\gamma/(2 \pi)$ &  ${\rm g}_{\rm n}$  \\
              &  (\%)    &       & ( JT$^{-1}$)           & (MHzT$^{-1}$)            &    \\
    \hline
    $^{89}$Y   & 100  & 1/2   & -0.13742 & -2.09 & -0.2748 \\
    \hline
    $^{28}$Si  &92.23 &  0  &    & 0 &    \\
    $^{29}$Si  & 4.68 &1/2 & -0.55529 &  -8.46 & -1.11058 \\
    $^{30}$Si  & 3.09 &  0  &    & 0 &    \\
    \hline
    $^{16}$O   &99.75 &  0  &    & 0 &    \\
    $^{17}$O   &0.038 &5/2 & -1.89380 &  -5.77 & -0.75752 \\
    $^{18}$O   &0.205 &  0  &    & 0 &    \\
    \hline
    \hline
    $^{162}$Er   & 0.14  & 0   &  & 0 &  \\
    $^{164}$Er   & 1.61  & 0   &  & 0 &  \\
    $^{166}$Er   & 33.61  & 0   &  & 0 &  \\
    $^{167}$Er   & 22.93  & 7/2   & -0.5665 & -1.23 & -0.1618  \\
    $^{168}$Er   & 26.78  & 0   &  & 0 &  \\
    $^{170}$Er   & 14.93  & 0   &  & 0 &  \\
    \hline
  \end{tabular}
 \end{center}
 \label{TBL1A}
\end{table}


\section{Used spin Hamiltonian parameters} \label{AQgMatrices}

Here we show the used effective spin Hamiltonian parameters $\bm A$, $\bm Q$, and ${\rm {\mathbf g}}_{\rm e}$ tensors of the ground Stark level $^4 I_{15/2} \ (Z_1)$ of $^{167}$Er$^{3+}$:YSO, which were recently refined based on electron-paramagnetic-resonance~\cite{chen2018hyperfine} and Raman-heterodyne~\cite{rakonjac2020long} experiments by S-J. Wang \textit{et al.}~\cite{wang2023hyperfine}. They discussed and compared these parameters with the previously reported ones in details~\cite{guillot2006hyperfine,sun2008magnetic,chen2018hyperfine,horvath2019extending,jobbitt2021prediction}. 
The following parameters are matrices in optical frame $(D_1, D_2, b)$ and the unit of $\bm A$ and  $\bm Q$ tensors is MHz.
\begin{table}[htb]
 \caption{$\bm A$, $\bm Q$, and ${\rm {\mathbf g}}_{\rm e}$ tensors of the ground Stark level $^4 I_{15/2} \ (Z_1)$ of $^{167}$Er$^{3+}$:Y$_2$SiO$_5$~\cite{wang2023hyperfine}.}
 \begin{center}
  {
  \begin{tabular}{|c|c|c|}
    \hline
       & site 1    & site 2\\
    \hline
   $\bm A$ (MHz) & $\displaystyle
   \begin{pmatrix}
308& -275 & -273 \\
-275 & 821 & 716 \\
-273 & 716 & 569 \\
\end{pmatrix}$ & $\displaystyle
   \begin{pmatrix}
-1570& 224 & -131 \\
224 & -17 & -15 \\
-131 & -15 & 141 \\
\end{pmatrix}$ \\
    \hline
   $\bm Q$ (MHz) & $\displaystyle
   \begin{pmatrix}
9.3& -9.9 & -14.0 \\
-9.9 & -5.7 & -15.5 \\
-14.0 & -15.5 & -3.6 \\
\end{pmatrix}$ & $\displaystyle
   \begin{pmatrix}
-9.8& -21.0 & -0.4 \\
-21.0 & -16.0 & -12.4 \\
-0.4 & -12.4 & 25.8 \\
\end{pmatrix}$ \\
    \hline
   ${\rm {\mathbf g}}_{\rm e}$ & $\displaystyle
   \begin{pmatrix}
2.75& -2.91 & -3.52 \\
-2.91 & 8.98 & 5.69 \\
-3.52 & 5.69 & 5.11 \\
\end{pmatrix}$ & $\displaystyle
   \begin{pmatrix}
14.44& -1.76 &  2.35 \\
-1.76 & 1.91 & -0.46 \\
2.35 & -0.46 & 1.424 \\
\end{pmatrix}$ \\
    \hline
  \end{tabular} \label{TBLA1}
}
 \end{center}
\end{table}

\section{Evaluation of the distance between Er ions as a function of Er concentration} \label{A2}

\begin{figure}[htb]
    \centering
	\includegraphics[width=\columnwidth, clip]{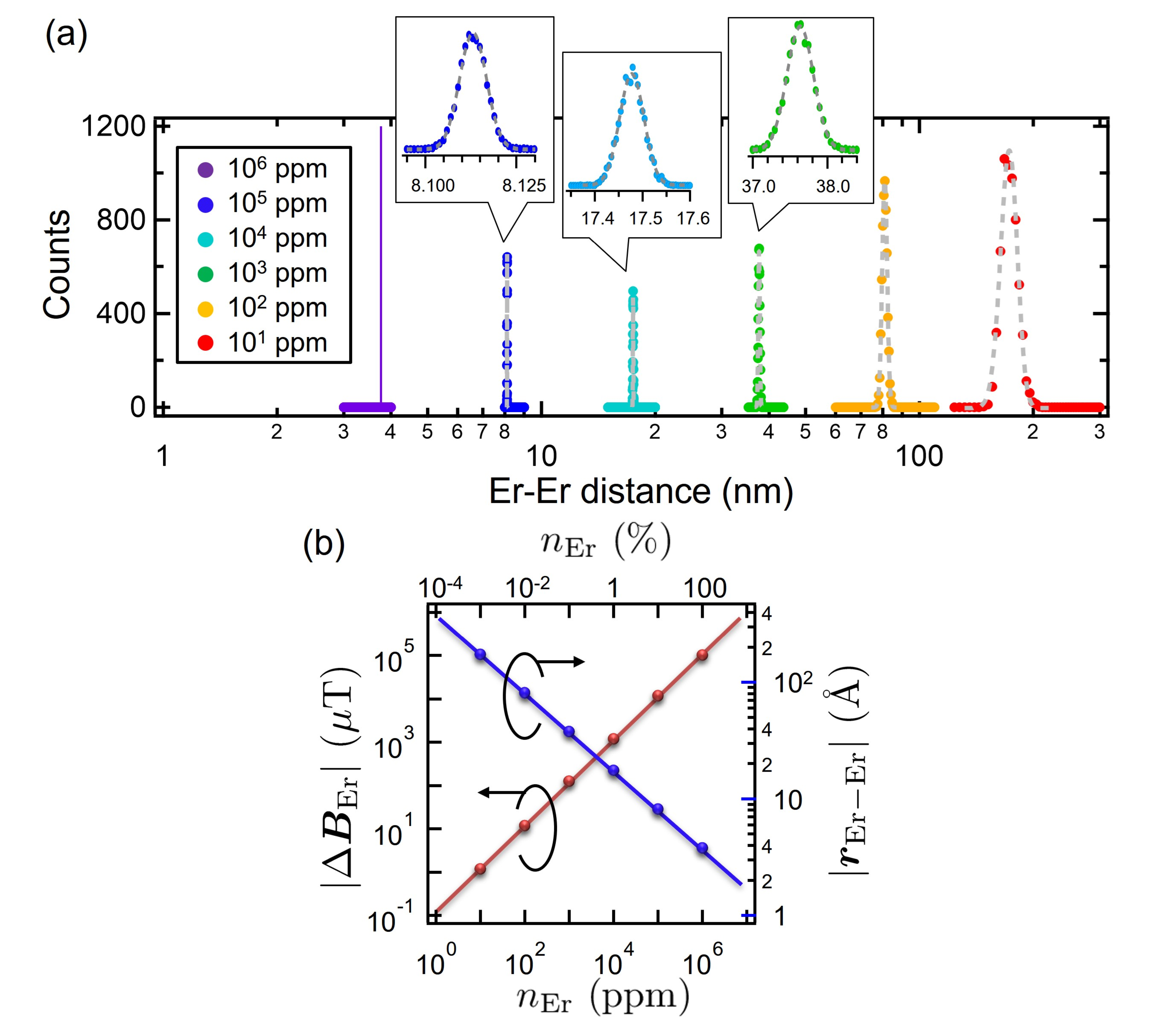}    
	\caption{(a) Distribution of inter-Er-ion distance $|\bm r_{\rm Er-Er}|$ and the dependence on $n_{\rm Er}$ calculated by Monte-Carlo method. (b) Dependence of $|\Delta \bm B_{\rm Er}|$ (red solid circles) and $|\bm r_{\rm Er-Er}|$ (blue solid circles) on $n_{\rm Er}$ obtained by Monte-Carlo method. The solid lines are based on Eqs. \eqref{eq9a} and \eqref{eq9c}.}
    \label{FigA2}
\end{figure}

The computed distribution of the distance between Er ions $|\bm r_{\rm Er-Er}|$ in Er$^{3+}$:YSO crystal and the dependence on $n_{\rm Er}$ (10-10$^6$ ppm) are shown in Fig.~\ref{FigA2}(a). The computed conditions by Monte-Carlo method are the same as those in Fig.~\ref{Fig2}. 

The $|\bm r_{\rm Er-Er}|$ can be obtained also by dividing the volume $V$ by the actual number of substituted  Er ions. The formula for the calculation is as follows:
\begin{subequations}
\begin{align}
|\bm r_{\rm Er-Er}| &= \left( \frac{V}{n P_{\rm Er}} \right)^{1/3}, \label{eq9a} \\
P_{\rm Er} &= \frac{2 m_{\rm Y}+m_{\rm Si}+5 m_{\rm O} }{ m_{\rm Er} }M_{\rm Er}, \label{eq9b}
\end{align}
\end{subequations}
where $P_{\rm Er}\ (M_{\rm Er})$ is Er concentration in number of ions (in weight), $m_{\rm A(=Y, Si, O)}$ is ion mass, $V$ is unit cell volume, and $n$ is number of Y ions in unit cell (=16). 
The unit cell of Y$_2$SiO$_5$ crystal has $a = 14.411$\AA, $b = 6.726 $\AA, $c = 10.419 $\AA, $\beta = 122.2^{\circ}$ and its volume is $V = abc \sin(\beta) = 854.57 \times 10^{-30}$ m$^3$~\cite{materialsproject}. One unit cell contains 16 Y ions, 8 Si ions, and 40 O ions. 

In Eq.~\eqref{eq6}, by setting the polar angle between the spins $\theta = 0$ and disregarding the anisotropy of the ${\rm {\mathbf g}}_{\rm e}$ tensor, we replace ${\rm {\mathbf g}}_{\rm e}$ with ${\rm g}_{\rm e}^{\rm eff} = 14.7$ \cite{sun2008magnetic} and $ \mu_{\rm{Er}} = \mu_{\rm B}{\rm g}_{\rm e}^{\rm eff}/2$ , leading to the following equation. 
\begin{subequations}
\begin{align}
|\Delta \bm B_{\rm{Er}}| = \frac{2 \mu_0 \mu_{\rm B}{\rm g}_{\rm e}^{\rm eff} }{4\pi |\bm r|^3} \label{eq9c}
\end{align}
\end{subequations}

In Fig.~\ref{FigA2}(b), the results from the Monte-Carlo method (circles) are compared with the analytical models of Eqs.~\eqref{eq9a}, ~\eqref{eq9b}, and ~\eqref{eq9c} (solid lines), showing excellent agreement across the plotted range. 
However, it is important to note that our analysis is based on dipole-dipole interactions between individual Er ions. This is a approximation that is to be effective in dilute systems. This is because, as the concentration increases significantly and approaches a fully substituted system (e.g., Er$_2$SiO$_5$), effects emerge that cannot be captured by this model, such as collective magnetic effects leading to ferromagnetic or antiferromagnetic ordering.
Therefore, while our model accurately describes the system in the concentration range relevant to this work, deviations would be expected at extremely high concentrations where the simple dipole-dipole picture is no longer sufficient. This validates the use of Eq. (C2a) to model the relationship between $n_{\rm Er}$ and $|\Delta \bm B_{\rm{Er}}|$ for our main analysis.

\section{Calculation method of $\bm S_1$ and $\bm S_2$} \label{A3}
\begin{figure}[htb]
  \centering
    \includegraphics[width=0.6\columnwidth, clip]{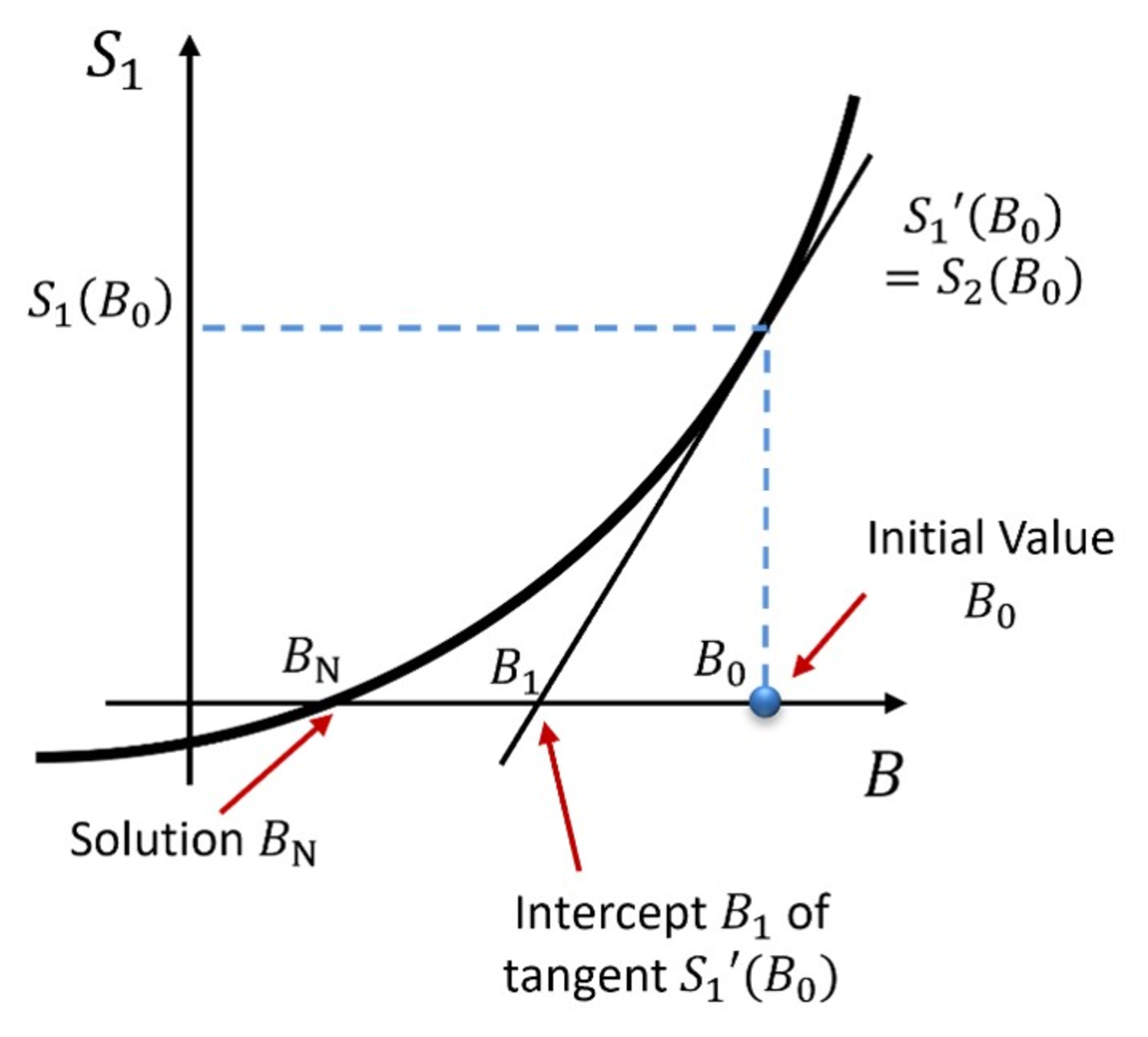}
\caption{Conceptual diagram of the Newton method. In the Newton method, a tangent is drawn from a certain value, and a new point is found where it intersects the y-axis. The
process is repeated by finding the tangent at the new point until the solution is reached.
}
\label{FigA3}
\end{figure}
In the Newton-Raphson method search for the magnetic fields where $\bm S_1(\bm B)$ approaches 0 by the following equation, 
\begin{equation*}
\bm B_{n+1}=\bm B_n-\frac{1}{2}\frac{\bm S_1}{\bm S_2}.
\end{equation*}
As shown in Fig.~\ref{FigA3}, 
\begin{enumerate}
  \item Draw the tangent line to $\bm S_1(\bm B_0)$ from an initial value $\bm B_0$.
  \item Use the intercept of this tangent as the next reference value $\bm B_1$.
  \item Draw the tangent line to $\bm S_1(\bm B_1)$.
  \item Repeat this process until $\bm S_1(\bm B_{\rm N})$ approaches 0.
\end{enumerate}
In this study, because the total number of ZEFOZ points found depends on the initial grid conditions, the initial magnetic fields for the search were established by combining the three conditions below:
\begin{itemize}
 \item A three-dimensional grid with coordinates ($B_{\rm D_1}, B_{\rm D_2}, B_{\rm b}$), where each component was chosen from the set {-25, -15, -5, 5, 15, 25} mT.
 \item A search was performed from 0 to 0.2 T with 8.6956 mT increments.
 \item A search was performed from 0.2 to 20 T with 1 T increments.
\end{itemize}
The number of iterations is limited to a maximum of 230 for computational convenience. We identified points among those that converged, selecting ZEFOZ points where $|S_1| < 10^{-8}$ Hz/T and the magnetic field $B < 10$ T.

\section{Discrepancies between experimental and calculated values near zero magnetic field.} \label{A4}

\begin{figure}[ht]
  \centering
    \includegraphics[width=1.0\columnwidth, clip]{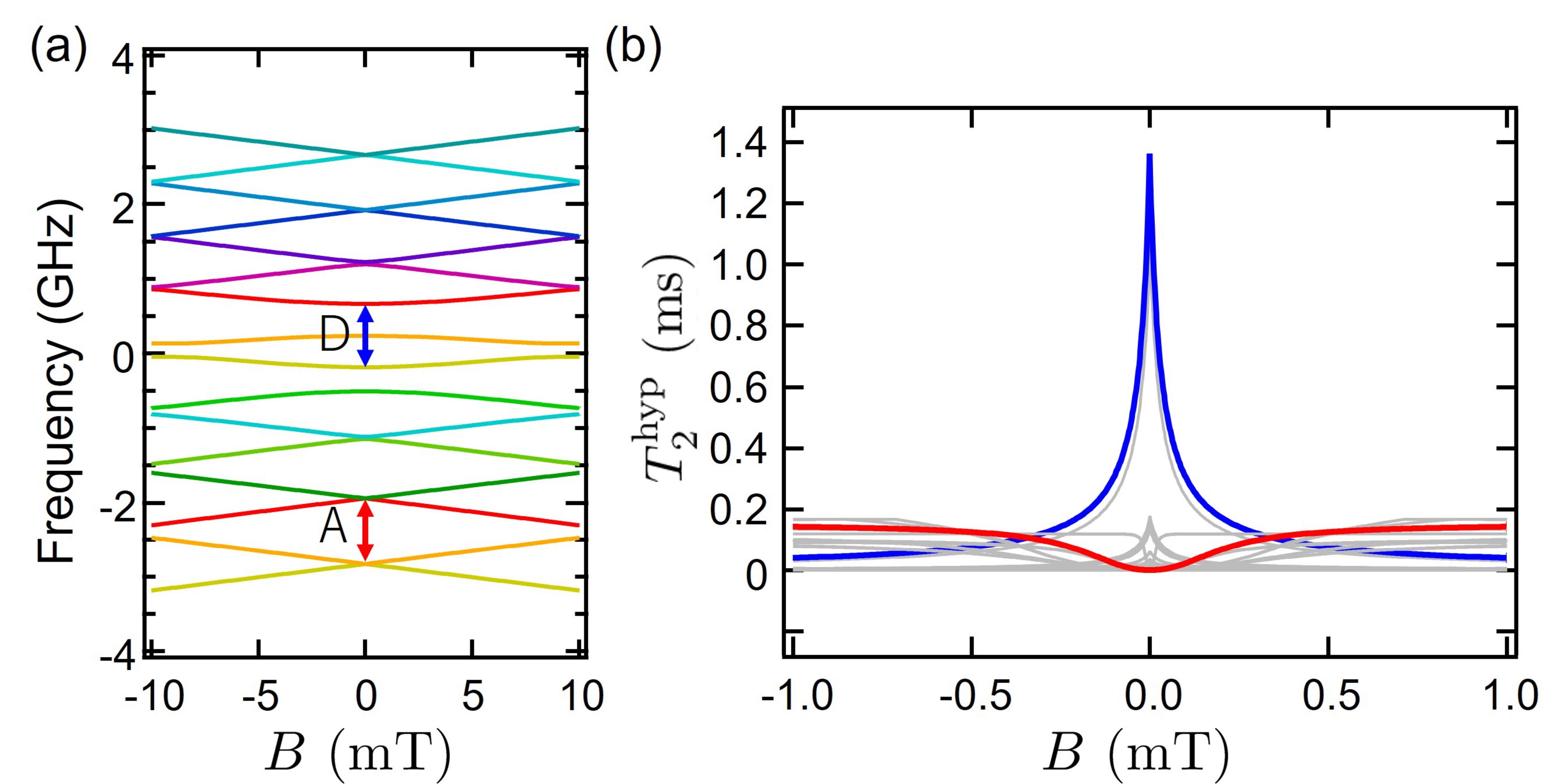}
\caption{(a) Energy structure of the ground state $^4 I_{15/2} \ (Z_1)$ for site 1 near zero magnetic field applied along $D_1$ axis. Two transitions A and D are indicated by the red and blue double-headed arrows, respectively. (b) The change in spin coherence time $T_2^{\rm hyp}$ with respect to applied magnetic field range of $-1 \sim 1$ mT. 
The blue (red) line represents $T_2^{\rm hyp}$ for transition D (A), and the gray lines represent $T_2^{\rm hyp}$ for the other transitions.
}
\label{FigA4}
\end{figure}

In previous study by J. V. Rakonjac \textit{et al.}~\cite{rakonjac2020long}, the two transitions labeled ''A'' and ''D'' for site 1 in Fig.~\ref{FigA4}(a) have the following HF levels, transition frequencies (measured values), and spin coherence times (measured values):
\begin{itemize}
    \item A: $|0 \rangle \Leftrightarrow |2 \rangle$, 880 MHz, $T_2^{\rm hyp} = 67 \, {\rm \mu s}$ (without DD)
    \item D: $|7 \rangle \Leftrightarrow |9 \rangle$, 774 MHz, $T_2^{\rm hyp} = 18 \, {\rm \mu s}$ (without DD)
\end{itemize}

As can be seen in Fig.~\ref{FigA4}(a), transition D is a ZEFOZ transition and transition A is called a double-doublet transition. 
They used 50-ppm $^{167}$Er$^{3+}$:YSO. In their work, $T_2^{\rm hyp}$ of their selected ZEFOZ transition D was shorter than that of
the doublet-doublet transition A. The reasons for this have not been elucidated in the paper. 

To investigate this discrepancy, we calculated the $T_2^{\rm hyp}$ for both transitions as a function of a small magnetic field applied along the $D_1$ axis, with the results shown in Fig.~\ref{FigA4}(b). Our calculations confirm that at an exact zero field, the ZEFOZ transition D has a coherence time significantly longer than that of the double-doublet transition A, as expected from theory. However, the coherence of transition D is highly sensitive to the magnetic field, decreasing rapidly and falling below that of transition A for fields greater than 0.3 mT. By comparing our calculations with the experimental data, we find that the observed ratio of spin coherence times ($T_2^{\rm hyp}$(A)/$T_2^{\rm hyp}$(D) $\approx$ 3.7) is well reproduced at a magnetic field of approximately 1 mT (see Table~\ref{TBLA3}). This suggests that a small, uncompensated stray magnetic field of this magnitude is a plausible explanation for the counter-intuitive results reported in the experiment.

\begin{table}[htb]
    \centering
    \tblcaption{Relationship between magnetic field $B$ and spin coherence time $T_2^{\rm hyp}$ for the transitions A and D.}
    \label{TBLA3}
    \begin{tabular}{|c|c|c|}
    \hline
    $B$ & $T_2^{\rm hyp}$(A) & $T_2^{\rm hyp}$(D) \\
    \hline
    0 mT   & 3 $\mu$s     & 1.3 ms   \\
    0.1 mT & 21 $\mu$s    & 305 $\mu$s   \\
    0.3 mT & 96 $\mu$s    & 119 $\mu$s   \\
    1.0 mT & 134 $\mu$s   & 38 $\mu$s    \\
    \hline
    Experimental & 67 $\mu$s    & 18 $\mu$s    \\
    \hline
    \end{tabular}
\end{table}

\section{Transition strength and frequency for some HF transitions with long spin coherence time} \label{A5}

Here we show five and ten transitions with long spin coherence time $T_2^{\rm hyp}$ at site 1 and site 2 in 10-ppm $^{167}$Er$^{3+}$:YSO under zero (Table~\ref{TBLA5-1}) and non-zero (Table~\ref{TBLA5-2}) magnetic fields.
\begin{table}[ht]
 \caption{Five transitions with long $T_2^{\rm hyp}$ at site 1 and site 2 under zero magnetic field in $^{167}$Er$^{3+}$:YSO with $n_{\rm Er}=$10 ppm. These transitions are mapped in Figs.~\ref{Fig3}(c) and~\ref{Fig3}(d), respectively.}
  \label{TBLA5-1}
 \begin{center}
  \begin{tabular}{|c|c|c|c|c|}
    \hline
       \multicolumn{5}{|c|}{site 1}             \\
    \hline
    No.  & Transition  & $T_2^{\rm hyp}$ & Strength   &  Frequency  \\
       &             &    ($\mu$s)        &  (GHz T$^{-1}$)  &  (MHz)     \\
    \hline
   1    & $\ket{7} \rightleftarrows \ket{9}$ & 4120 &9.1& 850.9  \\
    \hline
   2    & $\ket{6} \rightleftarrows \ket{8}$ & 3239 &7.3& 746.6 \\
    \hline
   3    & $\ket{8} \rightleftarrows \ket{9}$ & 507 &56.5& 427.6 \\
    \hline
   4    & $\ket{7} \rightleftarrows \ket{8}$ & 452 & 0.46 & 423.3\\
    \hline
   5    & $\ket{6} \rightleftarrows \ket{9}$ & 438 & 0.87 & 1174.2 \\
    \hline
  \end{tabular}
 \end{center}

  \begin{center}
  \begin{tabular}{|c|c|c|c|c|}
    \hline
       \multicolumn{5}{|c|}{site 2}             \\
    \hline
    No.  & Transition  & $T_2^{\rm hyp}$ & Strength   &  Frequency  \\
       &             &    ($\mu$s)         &  (GHz T$^{-1}$)  &   (MHz)     \\
    \hline
   1    & $\ket{7} \rightleftarrows \ket{9}$ & 1058 & 7.4 & 915.9\\
    \hline
   2    & $\ket{6} \rightleftarrows \ket{8}$ & 972 &8.2& 743.2\\
    \hline
   3    & $\ket{8} \rightleftarrows \ket{9}$ & 403 &17.3&  377.9 \\
    \hline
   4    & $\ket{6} \rightleftarrows \ket{10}$ & 354 &1.9& 1703.6 \\
    \hline
   5    & $\ket{7} \rightleftarrows \ket{11}$ & 342 &1.8& 1588.4 \\
    \hline
  \end{tabular}
 \end{center}
\end{table}

\begin{table}[htb]
 \caption{Ten transitions with long $T_2^{\rm hyp}$ at site 1 and site 2 under non-zero magnetic field in $^{167}$Er$^{3+}$:YSO with $n_{\rm Er}=$10 ppm. The plus (minus) sign of $\theta$ represent the magnetic subsite of orientation 1 (orientation 2) for site 1. In the case for site 2, the sign is reversed. These transitions for orientation 1 are mapped in Figs.~\ref{Fig4}(a) and~\ref{Fig4}(b), respectively.}
 \label{TBLA5-2}
 \begin{center}
  \begin{tabular}{|c|c|c|c|c|c|}
    \hline
       \multicolumn{6}{|c|}{site 1 [orientation 1 (2): $\theta>0 \ (\theta<0)$]}             \\
    \hline
    No.   & Transition  & $T_2^{\rm hyp}$ & Strength   &  Frequency & $\bm B(B, \theta, \phi)$ \\
       &             &     (s)        &  (MHz T$^{-1}$)  &   (MHz)    &  (T, $^{\circ}$,$^{\circ}$) \\
    \hline
   1    &$\ket{10} \rightleftarrows \ket{11}$& 173.9 & 25.5  & 783.0 &2.062, $\pm$53.92, -30.67 \\
    \hline
   2    &$\ket{5} \rightleftarrows \ket{6}$& 114.8 & 14.9 & 736.3 &2.134, $\pm$35.64, -13.62 \\
    \hline
   3    &$\ket{9} \rightleftarrows \ket{10}$&94.8& 35.3 & 828.3 &0.959, $\pm$59.69, -38.49  \\
    \hline
   4    &$\ket{8} \rightleftarrows \ket{9}$&71.4& 35.5 & 885.3 &0.741, $\pm$61.53, -43.00 \\
    \hline
   5    &$\ket{9} \rightleftarrows \ket{11}$&61.5& 2.5 & 1611.2 &1.320, $\pm$56.94, -34.59 \\
    \hline
   6    &$\ket{9} \rightleftarrows \ket{12}$&60.4& 0.14 & 2363.5 &2.201, $\pm$53.45, -30.18 \\
    \hline
   7    &$\ket{6} \rightleftarrows \ket{7}$&49.1& 11.4 & 745.8 &1.998, $\pm$37.68, -16.59 \\
    \hline
   8    &$\ket{4} \rightleftarrows \ket{7}$&44.3& 0.18 & 2216.2 &2.475, $\pm$36.31, -14.16 \\
    \hline
   9    &$\ket{8} \rightleftarrows \ket{10}$&40.6& 1.98 & 1713.5 &0.831, $\pm$60.71, -40.79 \\
    \hline
   10    &$\ket{8} \rightleftarrows \ket{13}$&38.6& 0.00 & 3985.6 &2.451, $\pm$52.72, -29.41 \\
    \hline
  \end{tabular}
 \end{center}
 %
 %
  \begin{center}
  \begin{tabular}{|c|c|c|c|c|c|}
    \hline
       \multicolumn{6}{|c|}{site 2 [orientation 1 (2): $\theta<0 \ (\theta>0)$]}             \\
    \hline
    No.   & Transition  & $T_2^{\rm hyp}$ & Strength   &  Frequency & $\bm B(B, \theta, \phi)$ \\
       &             &     (s)        &  (MHz T$^{-1}$)  &   (MHz)    &  (T, $^{\circ}$,$^{\circ}$) \\
    \hline
   1    &$\ket{14} \rightleftarrows \ket{15}$& 95.3 & 33.7 & 796.3 &0.633, $\mp$37.53, -10.94\\
    \hline
   2    &$\ket{6} \rightleftarrows \ket{7}$& 89.5 & 32.4 & 797.6 &0.668, $\mp$38.16, -12.59 \\
    \hline
   3    &$\ket{13} \rightleftarrows \ket{14}$& 54.3 & 57.4 & 796.0 &0.413, $\mp$40.22, -11.85 \\
    \hline
   4    &$\ket{13} \rightleftarrows \ket{15}$& 36.0 & 3.30 & 1592.2 &0.510, $\mp$38.88, -11.42\\
    \hline
   5    &$\ket{12} \rightleftarrows \ket{13}$& 30.4 & 84.7 & 798.0 &0.283, $\mp$42.23, -12.58\\
    \hline
       6    &$\ket{5} \rightleftarrows \ket{7}$& 29.7 & 2.67 & 1594.6 &0.513, $\mp$39.34, -12.83\\
    \hline
   7    &$\ket{5} \rightleftarrows \ket{6}$& 28.1 & 75.2 & 797.0 &0.335, $\mp$40.78, -13.14\\
    \hline
   8    &$\ket{12} \rightleftarrows \ket{14}$& 20.6 & 6.26 & 1593.9 &0.344, $\mp$41.25, -12.24\\
    \hline
   9    &$\ket{12} \rightleftarrows \ket{15}$& 18.4 & 0.36 & 2390.0 &0.422, $\mp$40.03, -11.85\\
    \hline
   10    &$\ket{11} \rightleftarrows \ket{12}$& 14.0 & 125 & 803.2 &0.191, $\mp$43.64, -13.22\\
    \hline
  \end{tabular}
 \end{center}
\end{table}


\end{appendix}

\clearpage

\begin{thebibliography}{99}
\bibitem{comment1}$\bm Q$ includes not only pure quadrupole interaction but also pseudo-quadrupole interaction derived from second-order HFI.
\end{thebibliography}

\end{document}